\documentclass[12pt]{article}
\usepackage{jheppub}
\usepackage[utf8]{inputenc}
\usepackage{epsfig}
\usepackage{amsfonts}
\usepackage{amssymb}
\usepackage{enumitem}
\usepackage{amsthm}
\usepackage{subfig}
\usepackage{bm}
\usepackage{hyperref}
\usepackage{mathrsfs}
\usepackage{bbm}
\usepackage{cancel}
\usepackage{color}
\usepackage{accents}
\usepackage{relsize}
\usepackage{float, slashed, graphicx, amssymb}
\usepackage{shuffle}
\usepackage{tikz}

\usetikzlibrary{patterns,decorations.markings,decorations.pathmorphing,arrows.meta}
\tikzset{box/.pic={\filldraw[fill=black]  (0,0) circle (2.5pt);
				   \filldraw [fill=black] (0.5,0) circle (2.5pt);
			       \draw [line width=5pt] (0,0) -- (0.5,0);}}

\newcommand{\cgg}[1] {{ \textcolor{red}{#1}}}

\makeatletter
\newcommand \UPlus {\mathop {\operator@font \uplus }\limits }
\makeatother
\makeatletter
\newcommand \Bigcup {\mathop {\operator@font \bigcup }\limits }
\makeatother

\def\LabelNote#1{}
\def\Label#1{\label{#1}%
  \smash{\hbox to0pt{\raise1ex\hbox{\tiny[#1]}\hss}}}
\def\Cdot{{\cdot}}
  
\def\nn{\nonumber}

\def\eqn#1{eq.~(\ref{#1})}

\def\spa#1.#2{\left\langle#1\,#2\right\rangle}
\def\spb#1.#2{\left[#1\,#2\right]}

\def\be{\begin{equation}}
\def\ee{\end{equation}}
\def\bea{\begin{eqnarray}}
\def\eea{\end{eqnarray}}  
  
\title{On the kinematic algebra for BCJ numerators beyond the MHV sector}

\author[a,b,c]{Gang Chen,}
\author[b,d]{Henrik Johansson,}
\author[b]{Fei Teng,}
\author[b,e,f]{and Tianheng Wang}

\affiliation[a]{Department of Physics, Zhejiang Normal University, Jinhua, Zhejiang Province, China}
\affiliation[b]{Department of Physics and Astronomy, Uppsala University, Uppsala, Sweden}
\affiliation[c]{Centre for Research in String Theory, School of Physics and Astronomy, Queen Mary University of London, Mile End Road, London E1 4NS, U.K.}
\affiliation[d]{Nordita, Stockholm University and KTH Royal Institute of Technology,\\ Roslagstullsbacken 23, 10691 Stockholm, Sweden}
\affiliation[e]{Department of Physics, Nanjing University, Nanjing, Jiangsu Province, China}
\affiliation[f]{Institut f\"ur Physik und IRIS Adlershof, Humboldt-Universit\"at zu Berlin, Zum Gro{\ss}en Windkanal 6, 12489 Berlin, Germany}

\emailAdd{g.chen@qmul.ac.uk}
\emailAdd{henrik.johansson@physics.uu.se}
\emailAdd{fei.teng@physics.uu.se}
\emailAdd{tianheng.wang@physik.hu-berlin.de}

\preprint{UUITP-22/19 \\
\phantom{~} \hfill NORDITA 2019-064 \\
\phantom{~} \hfill HU-EP-19/17\\
\phantom{~} \hfill QMUL-PH-19-14}  

\abstract{The duality between color and kinematics present in scattering amplitudes of Yang-Mills theory strongly suggest the existence of a hidden kinematic Lie algebra that controls the gauge theory. While associated BCJ numerators are known on closed forms to any multiplicity at tree level, the kinematic algebra has only been partially explored for the simplest of four-dimensional amplitudes: up to the MHV sector. In this paper we introduce a framework that allows us to characterize the algebra beyond the MHV sector. This allows us to both constrain some of the ambiguities of the kinematic algebra, and better control the generalized gauge freedom that is associated with the BCJ numerators. 
Specifically, in this paper, we work in dimension-agnostic notation and determine the kinematic algebra valid up to certain ${\cal O}\big((\varepsilon_i \Cdot \varepsilon_j)^2\big)$ terms that in four dimensions compute the next-to-MHV sector involving two scalars. 
The kinematic algebra in this sector is simple, given that we introduce tensor currents that generalize standard Yang-Mills vector currents. These tensor currents controls the generalized gauge freedom, allowing us to generate multiple different versions of BCJ numerators from the same kinematic algebra. The framework should generalize to other sectors in Yang-Mills theory.}

\begin{document} 
\maketitle
\flushbottom
\section{Introduction}
In recent years we have seen a remarkable progress in our basic understanding of gauge and gravity theories by exposing new properties and structures of scattering amplitudes in these theories. One such property is the duality between color and kinematics, also known as Bern-Carrasco-Johansson (BCJ) duality~\cite{Bern:2008qj,Bern:2010ue}, which appears to control the perturbative structure of many different gauge theories~\cite{Bern:2008qj,Bern:2010ue,Bargheer:2012gv,Broedel:2012rc,Chiodaroli:2013upa,Bern:2013yya,Johansson:2014zca,Chiodaroli:2014xia,Johansson:2015oia,Chiodaroli:2015rdg,Johansson:2017srf,Chiodaroli:2018dbu} as well as some scalar effective theories~\cite{Chen:2013fya,Chen:2014dfa,Du:2016tbc,Cheung:2016prv,Carrasco:2016ldy,Mafra:2016mcc,Carrasco:2016ygv}. For pure Yang-Mills theory, the duality constrains the kinematic numerators of individual cubic Feynman-like diagrams to obey Jacobi relations, in close analogy to the color factors of the same diagrams. The color factors inherit their Jacobi relations from the Lie algebra of the gauge group, and the duality between color and kinematics suggests that there exists a hidden kinematic Lie algebra that gives rise to the kinematic numerators. If so, this implies that Yang-Mills theories are in general characterized by a pair of Lie algebras, one for color and one for kinematics. 

It is curious to note that the objects that the kinematic Lie algebra should compute, the kinematic BCJ numerators, are under better control than the algebra itself. Numerators that manifest the color-kinematics duality have been constructed at any multiplicity for tree-level amplitudes in pure Yang-Mills theory~\cite{BjerrumBohr:2010hn,Mafra:2011kj,Fu:2012uy,Mafra:2015vca, Bjerrum-Bohr:2016axv,Du:2017kpo,Chen:2017bug,Fu:2018hpu}.  For supersymmetric Yang-Mills theory, it has been observed that a recursive construction similar to Berends-Giele recursion~\cite{Berends:1987me} can be used to generate BCJ numerators after using appropriate non-linear gauge transformations~\cite{Mafra:2015vca}. Beyond tree level, many BCJ numerators are known for specific (multi-)loop amplitudes in various (super-)Yang-Mills theories~\cite{Bern:2010ue,Bern:2012uf,Bjerrum-Bohr:2013iza,Bern:2013yya,Nohle:2013bfa,Chiodaroli:2013upa,Johansson:2014zca,Chiodaroli:2014xia,Mogull:2015adi,He:2017spx,Johansson:2017bfl, Kalin:2018thp} and also for multi-loop form factors~\cite{Boels:2012ew,Yang:2016ear,Boels:2017ftb}. Other (more exotic) gauge theories~\cite{Bargheer:2012gv,Huang:2012wr,Huang:2013kca,Broedel:2012rc,Carrasco:2012ca,Chiodaroli:2014xia,Johansson:2015oia,Johansson:2017srf,Johansson:2018ues, Chiodaroli:2018dbu}, as well as string theories~\cite{Mafra:2011nw,Broedel:2013tta,Stieberger:2014hba,Huang:2016tag,Azevedo:2018dgo}, have been observed to obey color-kinematics duality, but in these cases the BCJ numerators are not known beyond the simplest amplitudes. 

The duality between color and kinematics appears to sit at the core of a number of hidden connections between different theories. The most prominent examples are the Kawai-Lewellen-Tye (KLT) relations~\cite{Kawai:1985xq,Bern:1998sv,BjerrumBohr:2010hn}, the BCJ double copy~\cite{Bern:2008qj,Bern:2010ue} and the Cachazo-He-Yuan (CHY) construction~\cite{Cachazo:2013gna,Cachazo:2013hca,Cachazo:2014xea}. These all relate gauge, gravity and effective theories in highly nontrivial ways, yet the relations have a simplicity that makes them useful tools for practical calculations. There exists other perspectives of the same underlying structures, such as the unifying relations of ref.~\cite{Cheung:2017ems}, Witten's twistor string~\cite{Witten:2003nn,Berkovits:2004jj} and various ambi-twistor string constructions~\cite{Cachazo:2012da,Mason:2013sva,Geyer:2014fka, Casali:2015vta,Geyer:2015bja,Geyer:2016wjx}. String theory is a constant source of new results that have a deep connection to color-kinematics duality~\cite{BjerrumBohr:2009rd,Stieberger:2009hq,BjerrumBohr:2010hn,Mafra:2011kj,Mafra:2011nw,Broedel:2013tta,Stieberger:2014hba,Stieberger:2016lng,Huang:2016tag,Tourkine:2016bak,Hohenegger:2017kqy,Mafra:2017ioj,Ochirov:2017jby,Azevedo:2018dgo, He:2018pol}. However, when it comes to the kinematic Lie algebra there is currently no satisfactory construction from string theory, or from other constructions (see e.g. ref.~\cite{Fu:2016plh}), that yields local Yang-Mills numerators from a closed algebra. For non-local numerators there exists an algebraic construction using string vertex operators~\cite{Fu:2018hpu}. 

To this date, the kinematic Lie algebra of Yang-Mills theory remains enigmatic. On general grounds it can be expected to be an infinite-dimensional Lie algebra, since it must involve continuous parameters in the form of momentum, but the details of the generators and structure constants are mostly unknown. The exception, as shown by Monteiro and O'Connell~\cite{Monteiro:2011pc}, is the self-dual sector of Yang-Mills theory, where one can realize a subalgebra. After writing the self-dual gauge field in terms of a scalar field, the generators can now be simply labeled by the momenta of the scalar exitations, and further, be explicitly expressed in terms of plane-wave factors and simple differential operators (see section~\ref{sec:OM}). Commuting two generators gives rise to a Lie algebra that respects momentum conservation, and that has structure constants in the precise form of a cubic Feynman rule. All the Feynman diagrams of the self-dual sector can be computed through the Lie algebra. 

When relaxing away from the self-dual sector, one should expect that it is not sufficient to label the generators by only momenta. Indeed, on general grounds one can argue that there should exist generators that are labeled by similar kinematic parameters and indices as the physical states (see e.g. ref.~\cite{Monteiro:2013rya}), otherwise one cannot hope to use the algebra to compute scattering amplitudes with such external states. However, without specialized gauge choices, even the numerators in the maximally-helicity-violating (MHV) sector could not be computed from a local kinematic algebra that makes use of only the physical states.  As anticipated in the original work~\cite{Bern:2008qj}, and shown in detail in ref.~\cite{Bern:2010yg}, it appears necessary to introduce auxiliary fields in order the make Feynman rules both cubic and manifestly obey color-kinematics duality. Such auxiliary fields amend the list of generators with new ones that do not correspond to physical fields.

In the construction of Cheung and Shen~\cite{Cheung:2016prv}, a cubic action was presented that can be used to compute scattering amplitudes that manifestly obey color-kinematics duality. While the original motivation was to use it to study pion amplitudes in the non-linear-sigma model (NLSM)~\cite{Kampf:2012fn,Kampf:2013vha}, the action can alternatively be used to compute certain MHV amplitudes in Yang-Mills theory~\cite{Cheung:2017ems, Cheung:2017yef}, where two of the gluons are identified as scalars through a dimensional reduction procedure. We note that the BCJ numerators for the full MHV sector can be obtained by superposing the numerators obtained from the Cheung and Shen Lagrangian following the dimensional-oxidation prescription of ref.~\cite{Chiodaroli:2017ngp}. Thus one can argue, with some reservations as to the indirectness of the construction, that the MHV sector of tree-level Yang-Mills theory is by now understood in terms of a kinematic Lie algebra. 


In this paper, by introducing tensor currents, we propose a new approach to realize the algebra in a particular next-to-MHV (NMHV) sector of the Yang-Mills amplitudes. We view vector and tensor currents as the abstract generators of the algebra and define the fusion product rules between them. We find that in this particular sector, the fusion products are fixed after imposing some mild assumptions as well as physical constraints, such as gauge invariance and absence of amplitudes with tensor states. The algebra enables us to write down a closed formula for BCJ numerators in this sector valid for any multiplicity. An important bonus of our construction is that the tensors non-trivially encode the generalized gauge freedom related through use of the Clifford algebra. The gauge equivalence of different numerators can be understood through a set generalized BCJ relations that we describe in detail. 


The paper is organized as follows: In section~\ref{sec:review}, we review the two examples of the kinematic algebra~\cite{Monteiro:2011pc,Cheung:2016prv}. In section~\ref{sec:AlgConstr}, we introduce the concept of tensor current and present the algebraic construction. The closed form of the BCJ numerators, as a direct result of the algebra, is presented in section~\ref{sec:closedform}. The tensor currents capture a subspace of the generalized gauge freedom resulted from BCJ relations. Detailed discussions on generating different versions of BCJ numerators from the tensor currents are given in section~\ref{sec:FreedomBCJ}. We demonstrate that these different versions are equivalent at the level of the amplitudes in section~\ref{sec:TopInv}.

\section{Review and previous constructions of kinematic algebra}\label{sec:review}

Here we briefly review the kinematic-algebra constructions of Monteiro and O'Connell for self-dual YM and of Cheung and Shen for the MHV sector. We clarify some details of these constructions, and summarize the status of the MHV sector kinematic algebra.   

\subsection{Color-kinematics duality at tree level}

In the context of tree-level pure Yang-Mills theory, color-kinematics duality refers to the statement that we can write tree-level $n$-point scattering amplitudes as a sum over $(2n-5)!!$ cubic graphs
\be
{\cal A}_{n}^{\rm tree}= g^{n-2}\sum_{i=1}^{(2n-5)!!}\frac{n_i c_i}{D_i}
\label{YMtree}
\ee
where the color factors $c_i$ and kinematic numerators $n_i$ obey the same general Lie algebraic identities. Here the $D_i$ refers to the propagator denominator of the $i$'th graph. The $c_i$, which are built out of contractions of $f^{abc}$ structure constants of some gauge group $G$ following the cubic graph connectivity, obey three-term relations inherited from the Jacobi identity,
\be
c_i+c_j+c_k=0  ~~\sim~~ f^{abe}f^{ecd}+f^{ace}f^{edb}+f^{ade}f^{ebc}=0\,,
\ee 
for certain triplets of graphs $(i,j,k)$. Correspondingly, the kinematic numerators, which are functions built out of polarization vectors and momenta $n_i=n_i(\varepsilon_j, p_j)$, can be made to obey kinematic Jacobi relations for the same triplet of graphs,
 \be
c_i+c_j+c_k=0  ~~\Rightarrow~~ n_i+n_j+n_k=0\,.
\ee 
That such numerators can be found is the nontrivial statement of the duality. 

Kinematic numerators computed from standard Feynman rules of Yang-Mills theory do not obey the duality. Even if the quartic four-point interaction between gluons are turned into cubic interactions with the help of some auxiliary field\footnote{Cubic interactions are obtained by the replacement $ {\rm Tr}([A_\mu,A_\nu]^2)\rightarrow -\frac{1}{2}(B^{\mu \nu})^2+{\rm Tr}([A_\mu,A_\nu]B^{\mu \nu})$.} the duality is not inherited from the Feynman rules beyond four points. Since individual Feynman diagrams are not gauge invariant this observation is not in contradiction with the duality. Indeed, the cubic-graph decomposition~(\ref{YMtree}) is not unique because of the Jacobi-identity constraints satisfied by the color factors. This implies that the numerators possess a shift freedom that we refer to as {\it generalized gauge freedom},
\be
n_i\rightarrow n_i + \Delta_i\,, ~~~~~~\text{where }~~~~~~\sum_{i=1}^{(2n-5)!!} \frac{\Delta_i c_i}{D_i}=0\,.
\ee
This includes the usual gauge transformations that leaves the amplitude invariant $\varepsilon^\mu(p) \rightarrow \varepsilon^\mu(p)+p^\mu$, and generalizes it by allowing for any functions $\Delta_i$ that leaves the amplitude invariant.  The color-kinematics duality imply that starting from some generic cubic-graph representation of the amplitude, one can find some generalized gauge transformation that gives kinematic numerators that obey the duality. 

At tree level it is convenient work with a basis of BCJ numerators. By going to the color-kinematics-dual version of a Del-Duca-Dixon-Maltoni (DDM) \cite{DelDuca:1999iql,DelDuca:1999rs} multiperipheral basis, all numerators can be expressed using $(n-2)!$ permutations of the following graph,
\be
n(\sigma) \equiv n\big(\sigma_1,\sigma_2,\sigma_3,\ldots,\sigma_{n{-}1},\sigma_n\big)
 \equiv n\Bigg(\!\begin{tikzpicture}[baseline={([yshift=-0.8ex]current bounding box.center)}]\tikzstyle{every node}=[font=\tiny]
    \draw [decorate,decoration={coil,segment length=4pt}] (2.5,0) node[above=0pt] {$\sigma_n$} -- (-0.5,0) node[above=0pt] {$\sigma_1$};
                \draw [decorate,decoration={coil,segment length=4pt}] (1.8,0) -- ++(0,0.6) node[above=0pt] {$~~\sigma_{n{-}1}$};
                        \draw [decorate,decoration={coil,segment length=4pt}] (1.21,0) -- ++(0,0.6) node[above=0pt] {$\cdots$};
    \draw [decorate,decoration={coil,segment length=4pt}] (0.66,0) -- ++(0,0.6) node[above=0pt] {$\sigma_3$};
    \draw [decorate,decoration={coil,segment length=4pt}] (0.12,0) -- ++(0,0.6) node[above=0pt] {$\sigma_2$};
  \end{tikzpicture}\!\Bigg)\,,
\ee
and typically we fix $\sigma_1=1$ and $\sigma_n=n$, which gives the basis with $(n-2)!$ elements. 
And the color-ordered tree amplitude is then a sum over these $(n-2)!$ numerators,
\be
A^{\rm tree}_n(\rho_1,\rho_2,\ldots,\rho_n) =\sum_{\sigma \in S_{n-2}} m(\rho | \sigma ) \, n(\sigma) 
\ee
where $m(\rho | \sigma )$ is called the propagator matrix~\cite{Vaman:2010ez}. It is the same matrix as describes doubly color-ordered amplitudes in the bi-adjoint $\phi^3$ theory~\cite{Du:2011js,BjerrumBohr:2012mg,Cachazo:2013iea,Arkani-Hamed:2017mur,Bahjat-Abbas:2018vgo}. This matrix is related (by a pseudoinverse) to the Kawai-Lewellen-Tye (KLT) matrix \cite{Kawai:1985xq,Bern:1998sv,BjerrumBohr:2010hn}.

\subsection{The Monteiro-O'Connell construction}\label{sec:OM}
The kinematic algebra underlying the color-kinematics duality was first systematically studied by Monteiro and O'Connell  in ref.~\cite{Monteiro:2011pc}. They concluded that the self-dual Yang-Mills sector (which only gives non-zero amplitudes for the all-plus-helicity sector at one loop~\cite{Boels:2013bi}) automatically obeys a kinematic algebra.  Interactions in this sector can be simply be obtained by truncating the standard YM Feynman rules. First the quartic contact interactions are thrown away, and then only the cubic vertex corresponding to helicity $(++-)$ is kept. To isolate this helicity configuration, it is convenient to work with non-covariant objects. 

It is well known that the self-dual spectrum of YM consists of only the plus helicity $A^+$ field, which can be expressed in terms of a scalar field $A^+=\partial_u\Phi$. The non-abelian scalar $\Phi= \Phi^a T^a$ obeys the field equation 
\begin{equation}
\Box \Phi = -i  g [\partial_w \Phi,  \partial_u \Phi] \,,
\label{EQM}
\end{equation}
where $u$ and $w$ are mutually-orthogonal null directions: $u^2=w^2=u\cdot w=0$. Tree-level Feynman diagrams can be generated by recursively solving the field equation (proper external states are obtained by the inverse field map $ \Phi=\frac{1}{\partial_u} A^+$). In practice, for generating Feynman diagrams, the only non-trivial information in \eqn{EQM} is the kinematic dependence of the non-linear interaction term. After transforming to momentum space, it is captured by the function\footnote{As pointed out in ref.~\cite{Monteiro:2011pc}, this can be thought of as a convenient off-shell generalization of the inner product of on-shell Weyl spinors $X(p_1, p_2) \sim \spb{1}.{2}= \bar \lambda_{p_1}^\alpha  \bar \lambda_{p_2}^\beta \varepsilon_{\alpha \beta}$. }
\be
X(p_1, p_2) =p_{1 w} p_{2 u}-p_{2 w} p_{1 u}\,.
\ee
Tree-level Feynman diagram numerators are now simply given by cubic diagrams where each vertex represents the scattering process $\Phi_1 \Phi_2 \rightarrow \Phi_3 $ or equivalently in terms of the gluons $  A^+_1 A^+_2 \rightarrow A^+_3$, and thus the vertex is correspondingly dressed by a $X(p_1, p_2)$ factor.

Alternatively, the kinematic numerator can be generated by a kinematic Lie algebra, whose generators are labeled by momenta~\cite{Monteiro:2011pc}
\be
L_p=e^{-i p\cdot x}(p_u \partial_w-p_w \partial_u)
\ee
and the commutator $[A,B]=AB-BA$ can now be computed 
\begin{equation}
[L_{p_1}, L_{p_2}] = i X(p_1, p_2)L_{p_1+p_2}
\end{equation}
From the algebra we see that $X(p_1, p_2)$ are the kinematic structure constants of self-dual YM.  
The numerator can now be computed to any multiplicity by using nested commutators. For example, the four-point $s$-channel numerator is given by
\begin{equation}
N_s = [[L_{p_1}, L_{p_2}], L_{p_3}] = - X(p_1, p_2) X(p_1+p_2, p_3) L_{p_1+p_2+p_3}
\end{equation}
It obeys the Jacobi relation after summing over the $s,t,u$-channels
\bea
0&=&N_s+N_t+N_u = [[L_{p_1}, L_{p_2}], L_{p_3}] + {\rm cyclic}(1,2,3) \nn \\
&=&  -  \Big( X(p_1, p_2) X(p_1+p_2, p_3)  +  {\rm cyclic}(1,2,3) \Big)  L_{p_1+p_2+p_3}
\eea
So far, the numerators still contains a generator corresponding to the fourth external leg; to remove it, we can introduce a formal trace 
\be
n_s = {\rm Tr}(N_s \, L_{p_4}) = X(p_1, p_2) X(p_1+p_2, p_3) \delta^4(p_1+p_2+p_3+p_4)\,,
\ee
where the trace is normalized to ${\rm Tr}(L_{p} \,L_{p'})= \delta^4(p+p')$. Note that we include a delta function to make the dependence on $p_4$ manifest, but for the remaining part of this paper we drop such delta functions and consider the numerators only a function of the first $(n-1)$ momenta of an $n$-point amplitude.

The corresponding color factors can, of course, be generated by the same procedure from the gauge group Lie algebra, with generators $T^a$ and defining commutation relation $[T^a,T^b] = i f^{abc} T^c $. For example the four-point s-channel color factor is given by 
\be
C_s= [[T^{a_1},T^{a_2}],T^{a_3}]= - f^{a_1 a_2 b}f^{b a_3 c} T^c
\ee
We obtain the proper color factor after removing the last generator by tracing it against the external generator, $c_s = {\rm Tr}(C_s\, T^{a_4})= f^{a_1 a_2 b}f^{b a_3 a_4}$, using the normalization  ${\rm Tr}(T^{a} T^{b})=\delta^{ab}$. 

Finally, we note that Monteiro and O'Connell generalized their construction by considering a kinematic algebra induced by the CHY framework~\cite{Monteiro:2013rya}. While this in principle gives a construction for non-local BCJ numerators in terms of the solutions to the scattering equations, from the perspective of the current paper we are mainly interested in local numerators where the algebraic properties are manifest. This brings us to the Cheung-Shen construction.

\subsection{The Cheung-Shen construction}

In ref~\cite{Cheung:2016prv}  Cheung and Shen introduced a cubic action that given appropriate external wave functions computed pion scattering amplitudes, equivalent to those of a SU($N$) non-linear-sigma model. Moreover, the Feynman diagrams obtained computed from their action manifest the duality between color and kinematics.    
We can write the Lagrangian in the following form 
\be
{\cal L}_{\rm CS}= Z^{a \mu} \Box X_\mu^a+ \frac{1}{2} Y^a \Box Y^a- g f^{abc} Z^{a \mu}   \big(Z^{b \nu}  X^{c}_{\mu \nu}+ Y^b \partial_\mu Y^c \big)\,,
\label{Cheung_Shen}
\ee
where $Z^{a \mu}$ and $X^{\mu a}$ are complex vector fields transforming in the adjoint, and
$Y^a$ is an adjoint scalar. The tensor field is defined as $X^{a}_{\mu \nu}= \partial_\mu X^{a}_{\nu}-  \partial_\nu X^{a}_{\mu}$\,, and pions correspond to having either $Y^a$ or $ \partial_\mu Z^{a \mu}$ as external fields. 

In this paper, we are not concerned with pion amplitudes, instead we are interested in tree-level Yang-Mills amplitudes. To that end we note that the Cheung-Shen Lagrangian can be used to compute certain terms in a Yang-Mills amplitude. In particular, we note that it correctly reproduces MHV amplitudes with two adjoint scalars and $(n-2)$ gluons
\be
A(Y_1, g_2^+, \ldots,g_{x-1}^+,g_x^-, g_{x+1}^+,  \ldots, g_{n-1}^+, Y_n )
\ee
where $Y_i$ are scalars and $g_i^\pm \sim Z_i^\pm$ denote gluons, and the polarization vectors of the gluons needs to be ``gauge fixed'' as
\be
\varepsilon_i^{+ \mu}(p_i,p_x)= \frac{\langle i | \sigma^\mu | x ]}{\sqrt{2} \spb{i}.{x}}\,,~~~~~~\varepsilon_x^{-\mu }(p_x,q)= \frac{\langle q | \sigma^\mu | x ]}{\sqrt{2}\spa{q}.{x}}\,,
\ee
where for the MHV configuration only gluon $x$ has negative helicity, and the reference null momenta $q$ is arbitrary. With this choice any product $(\varepsilon_i {\cdot} \varepsilon_j)$ will vanish, and one can check that the Lagrangian (\ref{Cheung_Shen}) gives all the correct terms of type $(\varepsilon_i {\cdot}p_j)$ in a Yang-Mills amplitude. 

Furthermore, we can formally think of the two scalars as having their origin as extra-dimensional gluons. With this interpretation we can stretch the validity of the Cheung-Shen Lagrangian to compute all terms in a pure-YM numerator that has the schematic form
\be
n(\sigma)\sim ( \varepsilon_1 {\cdot} \varepsilon_n) \, n^{\rm CS}(\sigma) \sim  (\varepsilon_1 {\cdot} \varepsilon_n) \prod_{i,j} (\varepsilon_i {\cdot} p_j)
\ee
While these terms are sufficient to correctly describe the MHV amplitude with two scalars, they are not sufficient for the pure-gluon MHV amplitude. This is because we are lacking terms of the type  $(\varepsilon_i {\cdot} \varepsilon_j) \prod (\varepsilon_k {\cdot} p_l)$, where $(i,j) \neq (1,n)$ are not fixed. However, such terms are in principle related to the known ones by appropriate relabeling of the external particles, i.e. by using crossing symmetry. 

We can restore the missing crossing symmetry of the numerator using the dimensional-oxidation prescription of ref.~\cite{Chiodaroli:2017ngp},
giving
\be
n(\sigma)=\hskip-0.4cm  \sum_{1\leqslant i<j \leqslant n} \hskip-0.2cm (\varepsilon_{i} \Cdot \varepsilon_{j}) \, n^{\rm CS}(\sigma)\Big|_{{\rm legs}\, i,j \rightarrow Y_i,Y_j\, {\rm scalars}}
\label{dim-oxid}
\ee
where the notation implies that we compute the Cheung-Shen numerator for a fixed multiperipheral ordering $\sigma$, but we consider all possible locations of the two scalar external states, which are summed over. Since the Cheung-Shen numerator obeys the kinematic Jacobi identity to any multiplicity, and these relations are linear in the numerators, it follows that the above superposition of numerators also obeys the same relations. Hence, we have, through the kinematic algebra defined by the Cheung-Shen Lagrangian, obtained local color-kinematics-satisfying pure-YM numerators that can be used to compute MHV amplitudes.

\subsection{Considerations for going beyond MHV}

In order to compute numerators that can be used to obtain NMHV amplitudes, further generalizations of the kinematic algebra are needed. In particular, in the Cheung-Shen numerators we are missing terms with more powers of $(\varepsilon_{i} \Cdot \varepsilon_{j})$. For future convenience, we classify individual terms in the numerators by the number of $(\varepsilon_i\Cdot\varepsilon_j)$ factors, which we call the \emph{polarization power}. It is useful to keep in mind that at tree level, the number of $s_{ij}$ factors is precisely one less than the polarization power.


We can work out a correspondence between four-dimensional helicity sectors and the terms of different polarization powers. 
Specifically, for a fixed helicity configuration, we choose the plus helicity polarization vectors to have the same reference momenta $q_+$, and the negative helicity ones have $q_-$. We further require $q_-$ be one of the momenta of the plus helicity legs and $q_+$ be one of the momenta of the minus helicity ones.
Then the only nonzero products between polarization vectors are $(\varepsilon_{i}^+ {\cdot} \varepsilon_j^{-})$, where $i$ and $j$ cannot be the special legs with momenta $q_\pm$. Hence for an N${}^{k}$MHV amplitude that has $(k+2)$ negative helicity gluons, the maximum number of nonzero factors $(\varepsilon_{i}^+ {\cdot} \varepsilon_j^{-})$ is precisely the polarization power $(k+1)$. Therefore, if we have the following numerator terms under control:
\begin{equation*}\setlength\arraycolsep{5pt}
	\begin{array}{ c c c c c }
		\text{polarization power one:} & (\varepsilon_{i} \Cdot \varepsilon_j) \prod (\varepsilon_k \Cdot p_l) \phantom{\Big|}
		& \longrightarrow  & \text{MHV} \\ 
		\text{polarization power two:} & (\varepsilon_{i_1} \Cdot \varepsilon_{j_1}) (\varepsilon_{i_2} \Cdot \varepsilon_{j_2}) (p_{i_3} \Cdot p_{j_3}) \prod (\varepsilon_k \Cdot p_l) \phantom{\Big|}
		& \longrightarrow & \text{NMHV} 
	\end{array} \,,
\end{equation*}
we can compute numerators that give correct MHV and NMHV amplitudes.

In this paper we will focus on constructing the kinematic algebra that generates the numerator contributions for the terms of type
\begin{equation}\label{eq:ss}
(\hat \varepsilon_1{\cdot}  \hat \varepsilon_n){\textstyle\prod} (\varepsilon_k {\cdot}p_l)~~~~{\rm and }~~~~(\hat \varepsilon_1 {\cdot} \hat \varepsilon_n) (\varepsilon_i {\cdot}\varepsilon_j) (p_m{\cdot}p_r)  {\textstyle\prod} (\varepsilon_k {\cdot} p_l) \,.
\end{equation}
We call this the \emph{bi-scalar} NMHV sector of Yang-Mills theory. The gluon polarizations $\hat \varepsilon_1,\hat \varepsilon_n$ corresponds to scalar modes that live in a higher-dimensional (internal) space which is orthogonal to spacetime, meaning that $\hat \varepsilon_i \Cdot \varepsilon_j=\hat \varepsilon_i \Cdot p_j=0 $. This restricts the numerators (and the amplitude) to terms proportional to $\hat \varepsilon_1\Cdot\hat \varepsilon_n$. We can in principle recover the numerators for pure-gluon amplitudes by a similar procedure as in \eqn{dim-oxid}, see ref.~\cite{Chiodaroli:2017ngp}.

\section{Abstract tensor currents and algebraic construction}\label{sec:AlgConstr}
In this section, we propose a novel framework for exploring the kinematic algebra behind the tree-level BCJ numerators in $D$-dimensional Yang-Mills theory.

We begin with the observation that the BCJ numerators of massless quark-gluon amplitudes can be related to those of pure gluon amplitudes by taking (some of) the quarks to be soft.  As an inviting example, we consider the tree-level scattering of three gluons with a quark-antiquark pair. In this simple case, the BCJ numerators of multiperipheral diagrams always contain a piece obtained directly by the Feynman rules for massless quarks, and an additional piece that is proportional to an inverse propagator. This additional piece vanishes in the limit that the quark momentum becomes soft, namely,

\begin{align}\label{eq:qqbar3g}
n \Bigg(\!\begin{tikzpicture}[baseline={([yshift=-0.8ex]current bounding box.center)}]\tikzstyle{every node}=[font=\tiny]
    \draw [decoration={markings, mark=at position 0.15 with \arrow{latex},mark=at position 0.42 with \arrow{latex},mark=at position 0.7 with \arrow{latex},mark=at position 0.92 with \arrow{latex}},postaction=decorate] (1.5,0) node[above=0pt] {$p_4$} -- (-0.5,0) node[above=0pt] {$q$};
    \draw [decorate,decoration={coil,segment length=4pt}] (1,0) -- ++(0,0.6) node[above=0pt] {$p_3$};
    \draw [decorate,decoration={coil,segment length=4pt}] (0.5,0) -- ++(0,0.6) node[above=0pt] {$p_2$};
    \draw [decorate,decoration={coil,segment length=4pt}] (0,0) -- ++(0,0.6) node[above=0pt] {$p_1$};
  \end{tikzpicture}\!\Bigg)
  &=  \bar{v}\slashed \varepsilon_{1}(\slashed p_{1}+\slashed q)\slashed\varepsilon_{2}  (\slashed p_{12}+\slashed q) \slashed\varepsilon_{3} u + \frac{1}{3} (p_1+q)^2  \, \bar{v}[\slashed\varepsilon_{1},\slashed\varepsilon_{2}] \slashed\varepsilon_{3} u  \nn \\
   &=  \bar{v} \slashed \varepsilon_{1}\slashed p_{1}\slashed \varepsilon_{2} \slashed p_{12} \slashed\varepsilon_{3} u + \mathcal O(q)\,,
\end{align}
where $p_{ij}=p_i+p_j$, and on the second line we retain only the $q$-independent terms.
For this quark-gluon scattering process, all the BCJ numerators can be expressed in terms of the multiperipheral ones~\eqref{eq:qqbar3g} via commutation relations. For example, the numerator for the following diagram is given by two consecutive commutators,
\begin{align}
n(123;\bar v u) \equiv &
~
n\Bigg(\!\!\begin{tikzpicture}[baseline={([yshift=-0.8ex]current bounding box.center)}]\tikzstyle{every node}=[font=\tiny]
  \draw [decoration={markings, mark=at position 0.7 with \arrow{latex}},postaction=decorate] (1.56,0.2) node[right=-2pt] {$p_4$} -- (1.2,0);
  \draw [decoration={markings, mark=at position 0.7 with \arrow{latex}},postaction=decorate] (1.2,0) -- (1.56,-0.2) node[right=-2pt] {$q$};
  \draw [decorate,decoration={coil,segment length=4pt}] (-0.6,0) node[below=0pt]{$p_1$} -- (0,0);
  \draw [decorate,decoration={coil,segment length=4pt}] (0,0) -- ++(0,0.6) node[above=0pt] {$p_2$};
  \draw [decorate,decoration={coil,segment length=4pt}] (0,0)  -- (0.6,0);
  \draw [decorate,decoration={coil,segment length=4pt}] (0.6,0)-- ++(0,0.6) node[above=0pt] {$p_3$};
  \draw [decorate,decoration={coil,segment length=4pt}] (0.6,0) -- (1.2,0);
  \end{tikzpicture}\!\!\Bigg)
  = 
  n \Bigg(\!\begin{tikzpicture}[baseline={([yshift=-0.8ex]current bounding box.center)}]\tikzstyle{every node}=[font=\tiny]
    \draw [decoration={markings, mark=at position 0.15 with \arrow{latex},mark=at position 0.42 with \arrow{latex},mark=at position 0.7 with \arrow{latex},mark=at position 0.92 with \arrow{latex}},postaction=decorate] (1.5,0) node[above=0pt] {$p_4$} -- (-0.5,0) node[above=0pt] {$q$};
    \draw [decorate,decoration={coil,segment length=4pt}] (1,0) -- ++(0,0.6) node[above=0pt] {$~ p_3 \Big]$};
    \draw [decorate,decoration={coil,segment length=4pt}] (0.5,0) -- ++(0,0.6) node[above=0pt] {$ ~ p_2 \Big] $};
    \draw [decorate,decoration={coil,segment length=4pt}] (0,0) -- ++(0,0.6) node[above=0pt] {$\Big[\Big[ p_1~$};
  \end{tikzpicture}\!\Bigg)  \\
  =
& ~ \bar{v}\slashed \varepsilon_{1}\slashed p_{1}\slashed\varepsilon_{2} \slashed p_{12}\slashed\varepsilon_{3} u - \bar{v}\slashed\varepsilon_{2}\slashed p_{2}\slashed \varepsilon_{1} \slashed p_{12}\slashed\varepsilon_{3} u 
 - \bar{v}\slashed\varepsilon_{3}\slashed p_{3}\slashed \varepsilon_{1} \slashed p_{13} \slashed\varepsilon_{2} u + \bar{v}\slashed\varepsilon_{3}\slashed p_{3}\slashed\varepsilon_{2} \slashed p_{23} \slashed \varepsilon_{1} u  + \mathcal O(q)\,, \nn
 \label{eq:BCJNumTra}
\end{align}
and the other independent numerator $n(132;\bar v u)$ is obtained from $n(123;\bar v u)$ by the relabeling $2\leftrightarrow 3$. 

We note that one can view $\bar v \slashed \varepsilon_i u$ and $\bar v \slashed p_i u$ as Lorentz contractions of $\varepsilon_i$ and $p_i$ with a vector $\bar v (q) \gamma^\mu u(p_4)  \propto \varepsilon^\mu(p_4,q)$. Up to normalization, it behaves as a gluon polarization vector of a fourth gluon with momentum $p_4$. Due to the Dirac equation of the involved spinors, it obeys
$\varepsilon \Cdot p_4= \varepsilon \Cdot q =0$,
where $q$ can now be interpreted as a reference momentum. 
More generally, quantities of the form $\bar v \slashed \varepsilon_i \cdots \slashed p_k\cdots \slashed\varepsilon_j \cdots \slashed p_l u$ are Lorentz contractions of $\varepsilon$'s and $p$'s with tensorial objects $\bar v\gamma^{\mu_i}\cdots\gamma^{\mu_k}\cdots\gamma^{\mu_j}\cdots\gamma^{\mu_l} u$. If we choose to antisymmetrize  them they will also obey the physical requirements of the polarization of a antisymmetric tensor field. However, for now we are interested in obtaining the vector contributions without antisymmetrizing the tensors.  Instead we can eliminate the tensor contractions, except for the canonically ordered one $\bar u \slashed{\varepsilon}_1\slashed{\varepsilon}_2\slashed{\varepsilon}_3 v$, by repeated use of the Clifford algebra $\{ \gamma^\mu, \gamma^\nu \} =2 \eta^{\mu \nu}$.  
Applying the replacement $\bar v 
\gamma_{\phantom{4}}^\mu u\rightarrow \varepsilon^\mu_4$ to the vectors (with $\varepsilon^\mu_4$ properly normalized), we get
\be
\label{eq:4pqcdbasis123}
\left.\begin{bmatrix}
\,n(123;\bar v u)\, \\ \,n(132;\bar v u)\,
\end{bmatrix}\right|_{q\rightarrow 0,\;\bar v \gamma_{\phantom{4}}^\mu u \rightarrow \varepsilon^\mu_4} 
= \begin{bmatrix}
\,n(1234)\, \\ \,n(1324)\,
\end{bmatrix}_\text{vector}+
\begin{bmatrix}
-s_{12}\bar v \slashed{\varepsilon}_1\slashed{\varepsilon}_2\slashed{\varepsilon}_3 u \\ s_{13}\bar v \slashed{\varepsilon}_1\slashed{\varepsilon}_2\slashed{\varepsilon}_3 u
\end{bmatrix}_\text{tensor}\,, 
\ee
where the Mandelstam invariants have the slightly unusual normalization $s_{i_1i_2\cdots i_r}=\frac{1}{2}(p_{i_1}+p_{i_2}+\cdots+p_{i_r})^2$, which we will adhere to in this paper. This expression contains the gluon BCJ numerators $n(\ldots)$, where the bi-spinor $\bar v \gamma_{\phantom{4}}^\mu u$ was swapped for vector $\varepsilon_4^\mu$, but it also contains an irreducible tensor contribution. While this contamination may seem problematic, we note that the tensor terms corresponds to generalized gauge transformations. 
Namely, these terms lie in the null space of the four-point propagator matrix $m(\sigma | \rho)$,
\begin{align}\label{eq:4pmab}
\begin{bmatrix}
\frac{1}{s_{12}}+\frac{1}{s_{23}} & -\frac{1}{s_{23}} \\ 
-\frac{1}{s_{23}} & \frac{1}{s_{13}}+\frac{1}{s_{23}}
\end{bmatrix}\begin{bmatrix}
-s_{12}\bar v \slashed{\varepsilon}_1\slashed{\varepsilon}_2\slashed{\varepsilon}_3 u\vphantom{\frac{1}{s_{12}}} \\ s_{13}\bar v \slashed{\varepsilon}_1\slashed{\varepsilon}_2\slashed{\varepsilon}_3 u\vphantom{\frac{1}{s_{12}}} 
\end{bmatrix}
=0\,.
\end{align}
Since the partial tree amplitudes are equal to the product between the propagator matrix and the BCJ numerators, it follows that the  tensors do not contribute to the four-point amplitude. Although such terms are unconventional, we are allowed to include them in the BCJ numerators.

We can choose to reduce all the tensors to $\bar v \slashed{\varepsilon}_1\slashed{\varepsilon}_3\slashed{\varepsilon}_2 u$ instead (or any linear combination of these two tensors). It is easy to check that regardless of the choice, the tensor terms are always in the null space of the four-point propagator matrix. Although the vector terms depend on the choices, the difference caused by the choices is also in the null space of the propagator matrix. The existence of the tensors provides a way to partly control the generalized gauge freedom of the BCJ numerators.

For higher multiplicity, we introduce the concepts of \emph{vector} and \emph{tensor currents} as a generalization of the vector and tensor contractions above. We then construct the fusion products of these currents such that a particular class of ordered fusion products give the BCJ numerators up to the NMHV order in the DDM basis. The BCJ numerators outside the DDM basis are all determined by the Jacobi identities as demanded by the color-kinematics duality. 


\subsection{Algebraic framework for kinematic algebra}\label{sec:framework}

Based on the experience with tensors at four points, we now generalize to a more abstract setting. The algebraic framework will consist of two building blocks, the \emph{vector/tensor currents} and their \emph{fusion products}. 
A tensor current $J^{(w)}_{\mathfrak a_1 \otimes \mathfrak a_2\otimes \cdots \otimes \mathfrak a_m }(p)$ is an abstract generalization of the tensor  $\bar v(q) \slashed {\mathfrak a}_1 \slashed {\mathfrak a}_2  \cdots\slashed {\mathfrak a}_m  u(p) $. The tensor current carries a momentum $p$ and labels $\mathfrak a_i\otimes \cdots \otimes \mathfrak a_j$, where $\mathfrak a_i$ can either be a polarization vector or a momentum.  It also carries a discrete label $w$ to distinguish different types of abstract currents with the same tensor label, which may behave differently in the fusion products. 
The rank of a tensor current counts the number of its components $\mathfrak a_i$. In particular, we call a rank-one tensor current $J^{(w)}_{\mathfrak a}(p)$ a \emph{vector current}. When $p$ is on-shell, namely, $p^2=0$ and $\slashed{p}u=0$, we use the following replacement to get back to the original meaning of the tensor current: 
\begin{equation}
J^{(w)}_{\mathfrak a_1 \otimes \mathfrak a_2\otimes \cdots \otimes \mathfrak a_m } (p) \rightarrow \bar v(q) \slashed {\mathfrak a}_1 \slashed {\mathfrak a}_2  \cdots\slashed {\mathfrak a}_m  u(p)\,.
\end{equation}
The replacement is independent of $w$, where and the spinor $\bar v$ carries soft momenta ($q\rightarrow 0$). This replacement takes place in the final step of our algebraic construction. The mass dimension of a current $J$ is one plus the number of momenta appearing in its tensor component. The current should behave linearly in the components
\begin{align}
  J^{(w)}_{\cdots \otimes (x\mathfrak a_i+y\mathfrak a'_i) \otimes \cdots} (p) & = xJ^{(w)}_{\cdots \otimes \mathfrak a_i \otimes \cdots} (p) + yJ^{(w)}_{\cdots \otimes \mathfrak a'_i \otimes \cdots} (p)\,,& &\text{for }x,y\in\mathbb{C}\,,
\end{align}
and more importantly, it should satisfy a relation that is equivalent to the Clifford algebra,
\begin{align}
  \label{eq:CliffordTensor}
  J^{(w)}_{\cdots \mathfrak a_{i} \otimes \mathfrak a_j \otimes \mathfrak a_{k} \otimes \mathfrak a_{l} \otimes \cdots} (p) + J^{(w)}_{\cdots \mathfrak a_{i} \otimes \mathfrak a_{k} \otimes \mathfrak a_j \otimes \mathfrak a_{l} \otimes \cdots} (p)
  = (2\mathfrak a_j \Cdot \mathfrak a_{k}) J^{(w)}_{\cdots \otimes \mathfrak a_{i}\otimes \mathfrak a_{l} \otimes \cdots } (p)\,.
\end{align}
We generalize an external gluon to a vector current $J_{\varepsilon_i}(p_i)$, where $\varepsilon_i$ denotes its polarization vector and $p_i$ its momentum. We assume this vector current is unique and thus drop the index $w$ for simplicity.\footnote{In Section~\ref{sec:YMSrules}, we also consider scalars. They will be viewed as gluons whose polarization vectors live in higher dimensions.}

A fusion product describes the interaction of tensor currents. We only consider two-to-one fusions, which is consistent with cubic interactions. 
From a physical perspective, these fusion products are the counterparts of the Feynman rules for cubic interaction vertices. The outcome of a fusion product is a (sum of) tensor current(s) weighted by appropriate structure constant(s), where the conservation of momentum is implied. While the Feynman rules give rise to the amplitudes, the fusion products give rise to the BCJ numerators.

\begin{figure}[t]
  \centering
  \begin{tikzpicture}[scale=1.6]
    \coordinate (O) at (0,0);
    \coordinate (A) at (135:1);
    \coordinate (B) at (135:1.5);
    \draw [decorate,decoration={coil,segment length=4pt}]  (O.center) -- ++(45:2) node[above=-1pt,font=\fontsize{12}{12}\selectfont] {$J_{\varepsilon_{\sigma_{n{-}1}}}$};
    \draw [dashed,very thick] (0,-0.5) -- (O.center) -- (A.center) -- ++(135:1) node[above=-1pt,font=\fontsize{12}{12}\selectfont] {$J_{\hat \varepsilon_1}\vphantom{J_{\sigma_2}}$};
    \draw [decorate,decoration={coil,segment length=4pt}] (B.center) -- ++(45:0.5) node[above=-1pt,font=\fontsize{12}{12}\selectfont] {$J_{\varepsilon_{\sigma_2}}$};
    \draw [decorate,decoration={coil,segment length=4pt}] (A.center) -- ++(45:1) node[above=-1pt,font=\fontsize{12}{12}\selectfont] {$J_{\varepsilon_{\sigma_3}}$};
    \path (O.center) -- (135:0.5) -- ++(45:1.5) node[above=-1pt] {$\cdots$};
  \end{tikzpicture}
  \caption{Fusion product in the canonical ordering. The algebra simplifies by demanding that $\hat \varepsilon_1$ correspond to a scalar, since the internal currents must carry this polarization.}
  \label{fig:fusionProduct}
\end{figure}
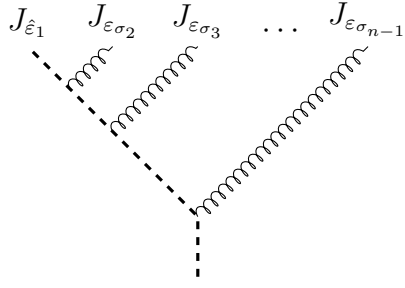

In general, we expect that the DDM basis numerator $N(\sigma_1,\sigma_2, \ldots, \sigma_{n{-}1},\sigma_n)\equiv N(\sigma)$ should be given by the kinematic analogue of a nested commutator. However, in the special bi-scalar sector, depicted in Figure~\ref{fig:fusionProduct}, the commutator structure should simplify to an ordered product of currents, where the pair-wise contractions are given by the fusion product, thus the numerator should take the form
\begin{align}
  N(\sigma) = J_{\hat \varepsilon_1}(p_1) \star J_{\varepsilon_{\sigma_2}}(p_{\sigma_2}) \star \cdots \star J_{\varepsilon_{\sigma_{n{-}1}}}(p_{\sigma_{n{-}1}})\,,
 \label{eq:current2num}
\end{align}
where as before $\sigma_1=1$ and $\sigma_n=n$.
We will interpret this product by evaluating it from left to right.  By construction the set of numerators are related by permutations in the legs $2$ to $n{-}1$. We will refer to such numerators as ``crossing symmetric'', even if leg $1$ and $n$ are always fixed and thus special.

Eq.~\eqref{eq:current2num} indicates that a fusion product is relevant to our construction only when its second input is a vector current $J_{\varepsilon_i}$. For a generic tensor current fused with a vector current, the fusion product is written as
\begin{align}
  \label{eq:tensorfusion}
J^{(w)}_{\mathfrak a_1\otimes \cdots \otimes \mathfrak a_m}(p) \star J_{\varepsilon_i}(p_i)=\sum_{k=1}^{m+2}\sum_{\mathfrak a'_1,\ldots, \mathfrak a'_k\,}\sum_{\,w'}f_{\mathfrak a_1\ldots \mathfrak a_m; \varepsilon_i}^{\mathfrak a'_1\ldots \mathfrak a'_k}(p,p_i;w,w')J_{\mathfrak a'_1\otimes\cdots\otimes \mathfrak a'_k}^{(w')}(p+p_i) \,,
\end{align}
where the $f$'s are coefficients that are similar to kinematic structure constants; although the  fusion product is not necessarily antisymmetric. Assuming that the $f$ coefficients are local polynomials in momenta and polarizations vectors gives the bound $k \leqslant m+2$. We note that we may think of the currents $J$ as being generators of the kinematic algebra, but we will not elaborate on that interpretation in this paper. 

In eq.~\eqref{eq:current2num}, $N(\sigma)$ is expressed in terms of vector and tensor currents. The tensor currents of the form $J_{\cdots\otimes p}(p)$ vanish when replaced by $\bar v \cdots \slashed p u$ because of the on-shell condition $\slashed p u=0$.
Using eq.~\eqref{eq:CliffordTensor}, we can bring the other tensor currents into an \emph{irreducible tensor current basis} $\mathsf b$, defined as the maximal set of tensor currents that cannot be linearly combined into vector currents or zero. This process generates additional vector currents.  A natural choice of $\mathsf{b}$ involves the tensor currents whose labels are in the ascending order:\footnote{The indices of the momenta have to be distinct because the on-shell condition $p_i^2=0$ implies that
$ J^{(w)}_{\cdots\otimes \mathfrak a\otimes p_i \otimes p_i \otimes  \mathfrak a'\otimes \cdots}(p) = 2 p_i^2  J^{(w)}_{\cdots\otimes \mathfrak a \otimes \mathfrak a' \otimes \cdots}(p)=0$ due to the relation~\eqref{eq:CliffordTensor}.}
\begin{align}\label{eq:CanonicalTensor}
J^{(w)}_{\varepsilon_{i_1} \otimes \varepsilon_{i_2} \otimes \cdots \otimes \varepsilon_{i_m}\otimes p_{j_1} \otimes p_{j_2} \otimes \cdots \otimes p_{j_{m'}} }(p)\,,& &\begin{array}{l}
1\leqslant i_1<\ldots <i_m\leqslant n-1 \\
1\leqslant j_1<\ldots <j_{m'}\leqslant n-1
\end{array} \,.
\end{align}
For tensor in the on-shell numerator $N(\sigma)$, we have $p=-p_n$ due to momentum conservation.
We can thus write eq.~\eqref{eq:current2num} as
\begin{align}
	N(\sigma)=N^{\mathsf T}_{\mathsf{b}}(\sigma)+N^{\mathsf  V}_{\mathsf b}(\sigma)\,,
	\label{eq:BCJNumTV}
\end{align}
where $N^{\mathsf T}_{\mathsf{b}}(\sigma)$ contains the terms spanned by the basis $\mathsf b$, and $N^{\mathsf  V}_{\mathsf b}(\sigma)$ denotes the corresponding vector current part. Both parts are affected by the choice of basis $\mathsf b$. While $N(\sigma)$ is crossing symmetric by construction, it is not necessarily true for $N^{\mathsf{T}}_{\mathsf b}(\sigma)$ and $N^{\mathsf{V}}_{\mathsf{b}}(\sigma)$ individually.
For the bi-scalar NMHV sector~\eqref{eq:ss}, a detailed discussion on this issue will be presented in section~\ref{sec:FreedomBCJ}.

For eq.~\eqref{eq:BCJNumTV} to be interpreted as the BCJ numerator associated with the DDM basis color ordering $\sigma$, the tensor and vector currents have to satisfy two conditions. These conditions will impose that only physical external states gives rise to non-vanishing amplitudes. Using the color-kinematics duality, the physical color-ordered amplitude is given by the action of the propagator matrix on the corresponding BCJ numerators. Therefore,  for unphysical states, under the action of the propagator matrix, the tensor part should vanish and the vector part should satisfy the gauge invariance condition, namely,  
\begin{align}
\bullet \,\,\,&\textit{null-space condition} \,\,\,\,\,\,\,\,&\sum_{\sigma\in S_{n-2}}m(\rho|\sigma)  N^{\mathsf T}_{\mathsf b}(\sigma)\,\,\,\,\,\,\,\,\,\,\,\,\,&= 0\,,\label{eq:CondTen}\\
\bullet \,\,\,&\textit{gauge invariance}\,\,\,\,\,\,\,\,\,\,\,\,&\sum_{\sigma\in S_{n-2}}m(\rho|\sigma) N^{\mathsf V}_{\mathsf b}(\sigma)\Big|_{\varepsilon_i\rightarrow p_i} &= 0\, .
\label{eq:CondVec}
\end{align}
where $m(\rho|\sigma)$ is the propagator matrix and $\rho,\sigma$ denote the color orderings.  The summation is taken over the DDM basis with the first and the last indices fixed. Fusion products satisfying these two constraints will give a kinematic algebra.

The standard BCJ numerator $n_{\mathsf b}(\sigma)$ can be read out from eq.~\eqref{eq:BCJNumTV} as follows
\begin{align}\label{eq:vecreplace}
n_{\mathsf b}(\sigma)= N^{\mathsf{V}}_\mathsf{b}(\sigma)\Big|_{J_{\mathfrak a}\rightarrow \mathfrak a\Cdot \hat \varepsilon_n}\,.
\end{align}
This lowers the dimension by one unit since the dimensionful $J_{\mathfrak a}$ gets replaced by a dimensionless polarization vector.  The subscript $\mathsf b$ emphasizes again that $N(\sigma)$ can be transformed into different versions of the BCJ numerators, depending on the choice of $\mathsf b$. The tensor part $N^{\mathsf T}_{\mathsf b}(\sigma)$ does not contribute to $n_{\mathsf b}(\sigma)$ as it satisfies the null-space condition above.

\subsection{Determining the fusion products for NMHV numerators}\label{sec:YMSrules}
As the main result of this paper, we derive the fusion products that give the BCJ numerators in the DDM basis for the bi-scalar NMHV sector~\eqref{eq:ss} of Yang-Mills theory. They can be solved for using an ansatz by imposing both the null-space condition~\eqref{eq:CondTen} and the gauge invariance condition~\eqref{eq:CondVec}, together with a few assumptions based on physical considerations.

In the NMHV sector, we may argue that a tensor current can only have up to three $\varepsilon$'s and two $p$'s, giving the following list of possible currents: 
\be\label{eq:currents}
J^{(w)}_{\varepsilon_i}\,,~~   J^{(w)}_{\varepsilon_i\otimes \varepsilon_j\otimes p_l}  \,, ~~
J^{(w)}_{\varepsilon_i \otimes \varepsilon_j \otimes \varepsilon_h}\,,~~J^{(w)}_{p_i}\,,~~
 J^{(w)}_{\varepsilon_i\otimes p_l\otimes p_m}\,,~~ J^{(w)}_{\varepsilon_i \otimes \varepsilon_j \otimes \varepsilon_h\otimes p_l\otimes p_m}\,,
\ee
where $J^{(w)}_{\varepsilon_i}$ denotes both the current $J_{\varepsilon_i}(p_i)$ associated with an external gluon and any possible internal vector current. More $\varepsilon_i$ factors lead to terms exceeding the polarization power-two limit when we permute the tensor components to reach a basis. Likewise, more $p_i$'s lead to structure constants of negative mass dimensions, and such nonlocal terms are excluded from our considerations.

Specializing to the bi-scalar part of the NMHV sector~\eqref{eq:ss} means that we are only interested in terms proportional to $\hat \varepsilon_1\Cdot\hat \varepsilon_n$. Such terms can only be generated if the intermediate currents carry the polarization $\hat \varepsilon_1$ (see Figure~\ref{fig:fusionProduct}). Thus, in this sector,  the vector currents $J^{(w)}_{p_i}$ and $J^{(w)}_{\varepsilon_{i >1}}$ are excluded. Similarly, all the higher-rank tensor currents without $\hat \varepsilon_1$ are excluded. For the remaining tensor currents, we note that since $\hat \varepsilon_1$ is formally an extra-dimensional object, we can always use the Clifford algebra~\eqref{eq:CliffordTensor} to freely move it to the first entry of the tensor. 

We can further simplify the problem by imposing the following constraints on the ansatz for fusion products:
\begin{enumerate}[label=(\arabic*)]
\item We assume there is only one tensor of the type $J^{(w)}_{\hat \varepsilon_1\otimes\varepsilon_i\otimes p_j}$, and thus we omit the superscript. 
We further assume that such a current containing an explicit momentum needs to be converted to a proper tensor when it is fused with a physical vector, and the relevant fusion product then becomes
$$
J_{\hat \varepsilon_{1}\otimes \varepsilon_{i}\otimes p_j }(p)\star J_{\varepsilon_k} (p_k)={1\over 2} (p \Cdot p_j) J^{(1)}_{\hat \varepsilon_{1}\otimes \varepsilon_{i}\otimes \varepsilon_k} (p+ p_k)\,.
$$
The above equation leads to\footnote{This fusion rule bears a similar structure to the factorization $\sum_{\text{s}}(\bar v\cdots \slashed p u)(\bar v \slashed\varepsilon_i u)=p^2\bar v\cdots \slashed\varepsilon_i u$, where the state sum $\sum_{\text{s}}\bar v u=\slashed p$ is used.}
\begin{align}\label{eq:tenRule1}
J_{\hat \varepsilon_{1}\otimes \varepsilon_{i}\otimes p }(p)\star J_{\varepsilon_k}(p_k)=\frac{1}{2}\, p^2 J^{(1)}_{\hat \varepsilon_{1}\otimes \varepsilon_{i}\otimes \varepsilon_k}(p+p_k)\,.
\end{align}
We find that $J_{\hat \varepsilon_1\otimes\varepsilon_i\otimes p}(p)$ is the only tensor of this type that appears in our construction. 
\item The structure constants of $J^{(w)}_{\hat \varepsilon_1\otimes \varepsilon_i\otimes \varepsilon_j}(p) \star J_{\varepsilon_k}(p_k)$ consists of only dot products of the form ${\varepsilon_I}\Cdot p$ or ${\varepsilon_I}\Cdot p_k$ , where $I\in \{i,j,k\}$.
\end{enumerate}

Let us briefly discuss what are the possible intermediate currents that are generated by the ordered fusion product~\eqref{eq:current2num}, given the above assumptions. As discussed in section~\ref{sec:framework}, the current $J_{\hat \varepsilon_{1}\otimes \varepsilon_{i}\otimes p}(p)$ vanishes on shell. Thus it is used solely as a vehicle to raise tensor rank in intermediate steps. The currents $J^{(w)}_{\hat \varepsilon_{1}\otimes \varepsilon_{i}\otimes \varepsilon_k}$ produced by eq.~\eqref{eq:tenRule1} can all be chosen to be of the same type, so we set $w=1$. By restriction to the bi-scalar sector the intermediate vector currents can only be $J_{\hat \varepsilon_1}$. 
Finally, the second assumption (\ref{eq:tenRule1}) implies that $J^{(w)}_{\hat \varepsilon_1\otimes \varepsilon_i\otimes \varepsilon_j} \star J_{\varepsilon_k}$ only produces rank-three currents of the form $J^{(w)}_{\hat \varepsilon_1\otimes \varepsilon_i\otimes \varepsilon_j}$. Hence, internal vector currents and $J_{\hat \varepsilon_1\otimes\varepsilon_i\otimes p}$ are exclusively generated by fusions of vector currents. 

Altogether, the fusion rules under the above assumptions do not generate the last two terms in the list \eqref{eq:currents}. In conculsion, only the following fusion products are relevant to our construction,
 \be\label{eq:relevantProd}
 J_{\hat \varepsilon_1} \star J_{\varepsilon_k} \,, ~~~
J^{(w)}_{\hat \varepsilon_1\otimes \varepsilon_i\otimes \varepsilon_j} \star J_{\varepsilon_k}\,, ~~~
J_{\hat \varepsilon_1\otimes \varepsilon_i \otimes p}\star J_{\varepsilon_k} \,,\nn
\ee
and by assumption the last one is given by eq.~\eqref{eq:tenRule1}.
It is now feasible to write down a generic ansatz for the fusion products and solve eq.~\eqref{eq:CondTen} and~\eqref{eq:CondVec} for the kinematic-dependent coefficients. 

We note that the gauge invariance condition~\eqref{eq:CondVec} concerns only the transformations $\varepsilon_i\rightarrow p_i$ with $2\leqslant i\leqslant n-1$. Furthermore,  a gauge transformation mixes terms with different polarization powers in the numerators (or equivalently, orders in Mandelstam variables). In the restricted sector under consideration, the condition~\eqref{eq:CondVec} should thus hold up to the first order in the Mandelstam variables. A full gauge invariance check requires higher order terms in the polarization power, which is beyond the scope of this study.

At multiplicity three, eq.~\eqref{eq:current2num} becomes $J_{\hat \varepsilon_1}(p_1)\star J_{\varepsilon_2}(p_2)$, whose vector part must reproduce the BCJ numerator $N^{\mathsf V}(123)=\varepsilon_2\Cdot p_1 J_{\hat \varepsilon_1}\rightarrow\hat \varepsilon_1\Cdot \hat \varepsilon_3\, \varepsilon_2\Cdot p_1$.  The tensor part $N^{\mathsf T}$ must vanish on shell, hence it must be proportional to $J_{\hat \varepsilon_{1}\otimes \varepsilon_{2}\otimes (p_1+p_2)}$.
At four points, the fusion product $J_{\hat \varepsilon_1}(p_1)\star J_{\varepsilon_2}(p_2)\star J_{\varepsilon_3}(p_3)$ needs to give the correct BCJ numerator for the four-point amplitude. Eq.~\eqref{eq:tenRule1} then fixes the coefficient of $J_{\hat \varepsilon_{1}\otimes \varepsilon_{2}\otimes (p_1+p_2)}$ in $J_{\hat \varepsilon_1}(p_1)\star J_{\varepsilon_2}(p_2)$, which leads to
\begin{align}\label{eq:vecRule}
J_{\hat \varepsilon_{1} }(p)\star J_{\varepsilon_{i} }(p_i) =  \varepsilon_{i}{\cdot} p\, J_{\hat \varepsilon_{1} }(p+p_i)- \frac{1}{2} J_{\hat \varepsilon_{1}\otimes \varepsilon_{i}\otimes (p+p_i) }(p+p_i)\,.
\end{align}
The tensor $J^{(1)}_{\hat \varepsilon_{1}\otimes \varepsilon_{2}\otimes\varepsilon_{3} }$ shows up in the result of $J_{\hat \varepsilon_1}(p_1)\star J_{\varepsilon_2}(p_2)\star J_{\varepsilon_3}(p_3)$. This current does not vanish on-shell. Instead, it belongs to the null space of the four-point propagator matrix $m(\gamma|\sigma)$. 

Starting from five points, we encounter the fusion products $J^{(w)}_{\hat \varepsilon_{1}\otimes \varepsilon_{i}\otimes \varepsilon_{j} }(p)\star J_{\varepsilon_{k} }(p_k)$ ($w=1$ for five points). As mentioned before, the fusion product depends on the index $w$ labeling the current type. In the ansatz for $J^{(w)}_{\hat \varepsilon_{1}\otimes \varepsilon_{i}\otimes \varepsilon_{j} }(p)\star J_{\varepsilon_{k} }(p_k)$, we assume each current in the resulting expression carries a different $w$-index. Imposing the conditions eq.~\eqref{eq:CondTen} and eq.~\eqref{eq:CondVec}, we obtain the fusion rule for $J^{(1)}_{\hat \varepsilon_{1}\otimes \varepsilon_{i}\otimes \varepsilon_{j} }(p)\star J_{\varepsilon_{k} }(p_k)$. 
 
At six points, we repeat the exercise and solve the constraints~\eqref{eq:CondTen} and~\eqref{eq:CondVec} for the undetermined $J^{(w)}_{\hat \varepsilon_{1}\otimes \varepsilon_{i}\otimes \varepsilon_{j} }(p)\star J_{\varepsilon_{k} }(p_k)$. We observe that some of the currents $J^{(w)}_{\hat \varepsilon_{1}\otimes \varepsilon_{i}\otimes \varepsilon_{j}  }$ are forced to follow the same fusion rule as $J^{(1)}_{\hat \varepsilon_{1}\otimes \varepsilon_{i}\otimes \varepsilon_{j} }$, while the rest must behave differently. That is, those following the same rule are identified as $J^{(1)}_{\hat \varepsilon_{1}\otimes \varepsilon_{i}\otimes \varepsilon_{j} }$ while the rest are a new kind of current, which we denote as $J^{(2)}_{\hat \varepsilon_{1}\otimes \varepsilon_{i}\otimes \varepsilon_{j} }$.  
 At seven points, we find that the currents generated by $J^{(2)}_{\hat \varepsilon_{1}\otimes \varepsilon_{i}\otimes \varepsilon_{j} } \star J_{\varepsilon_k}$ are also of the second kind. Hence, the algebra is closed under multiplication on the right by $J_{\varepsilon_k}$. And the final fusion rules are determined to be
\begin{align}\label{eq:tenRule2}
J^{(1)}_{\hat \varepsilon_{1}\otimes \varepsilon_{i}\otimes \varepsilon_{j} }(p)\star J_{\varepsilon_{k} }(p_k) & =  ( \varepsilon_{k}\Cdot p) J^{(2)}_{\hat \varepsilon_{1}\otimes \varepsilon_{i}\otimes \varepsilon_{j} }(p+p_k)\nonumber\\
&\quad + ( \varepsilon_{j}\Cdot p) J^{(1)}_{\hat \varepsilon_{1}\otimes \varepsilon_{i}\otimes \varepsilon_{k} }(p+p_k)-(\varepsilon_{i}\Cdot p) J^{(1)}_{\hat \varepsilon_{1}\otimes \varepsilon_{j}\otimes \varepsilon_{k} }(p+p_k) \,,\nonumber\\
 J^{(2)}_{\hat \varepsilon_{1}\otimes \varepsilon_{i}\otimes \varepsilon_{j} }(p)\star J_{\varepsilon_{k} }(p_k) &=( \varepsilon_{k} \Cdot p) J^{(2)}_{\hat \varepsilon_{1}\otimes \varepsilon_{i}\otimes \varepsilon_{j} }(p+p_k)\,.
\end{align}
Notice that the fusion product involving $J^{(1)}_{\hat \varepsilon_{1}\otimes \varepsilon_{i}\otimes \varepsilon_{k} }$ generates both the first and the second types of tensor currents. Interestingly, the type is correlated to the tensor labels of the currents. The current with the same tensor labels as the input one has to be of the second type while the rest are of the first type.

In summary, up to six points, we encounter new fusion products at each multiplicity, and at six points and beyond the algebra closes. Under our assumptions, the constraints~\eqref{eq:CondTen} and~\eqref{eq:CondVec} give a solution to the algebra without free parameters. The final fusion rules are given in eq.~\eqref{eq:tenRule1}, \eqref{eq:vecRule} and~\eqref{eq:tenRule2}, and only five types of currents are involved
\be
	J_{\hat \varepsilon_1}\,, ~~
		J_{\varepsilon_i}\,, ~~
	J_{\hat \varepsilon_1\otimes\varepsilon_i\otimes p}\,, ~~
	 J^{(1)}_{\hat \varepsilon_1\otimes\varepsilon_i\otimes \varepsilon_j}\,, ~~
J^{(2)}_{\hat \varepsilon_1\otimes\varepsilon_i\otimes \varepsilon_j}\,.
\ee
Moreover, we have observed that, in this bi-scalar NMHV sector, the algebra was determined up to one free parameter by the null-space condition~\eqref{eq:CondTen} at each multiplicity up to seven points. This free parameter can be viewed as the global normalization of the tensor current part of the BCJ numerator, with respect to the respective vector current part. The gauge invariance condition~\eqref{eq:CondVec} fixes this global normalization.

We further check the algebra by computing the BCJ numerators from eq.~\eqref{eq:current2num} to ten points. The constraints~\eqref{eq:CondTen} and~\eqref{eq:CondVec} are automatically satisfied, which strongly suggests that the prescription works for any multiplicity, and the algebra is a candidate for the kinematic algebra in the bi-scalar NMHV sector~\eqref{eq:ss}. 

Let us emphasize that the tensor currents of the form $J^{(w)}_{\hat \varepsilon_{1}\otimes \varepsilon_{i}\otimes\varepsilon_{j} }$ appear in the numerator at four and beyond. As we will elaborate in section~\ref{sec:FreedomBCJ}, the appearance of such tensor currents reflects the generalized gauge freedom of the BCJ numerators. 

We note that the above fusion products only compute the half-ladder diagrams in the DDM basis. Other diagrams are obtained by the Jacobi identities of the BCJ numerators. To compute BCJ numerators outside the DDM basis using the kinematic algebra, the fusion products have to be generalized and new fusion products need to be constructed in such a way that they are consistent with the kinematic Jacobi identities. We leave this work for the future.

\section{Closed form of the NMHV BCJ numerator }\label{sec:closedform}
From the algebra given in eq.~\eqref{eq:tenRule1}, \eqref{eq:vecRule} and~\eqref{eq:tenRule2},  we can directly compute the $n$-point bi-scalar NMHV BCJ numerators by evaluating the fusion product~\eqref{eq:current2num} from left to right. The BCJ numerators computed this way can be packaged into a closed all-multiplicity formula, which will be presented in this section.

As mentioned before, in the final expression of the BCJ numerator, we implicitly equate $J^{(w)}_{\mathfrak a_1\otimes \cdots \otimes \mathfrak a_r} $ to $ \bar v \slashed {\mathfrak a}_1\cdots \slashed {\mathfrak a}_r u$, regardless of the $w$-index, which can thus be omitted in our final result. The currents of the form $J_{\hat \varepsilon_1\otimes\varepsilon_i\otimes p}$ vanish due to $\slashed{p} u(p) = 0$.
Since our construction is crossing-symmetric in leg $2$ to $n{-}1$, it suffices to display the numerator associated with the color ordering $\{123\cdots n\}$, while the rest of the DDM basis can be obtained by index permutation. We write the result of~\eqref{eq:current2num} as
\be
N=N^{(1)}+N^{(2)}+\text{terms that vanish on shell}\,.
\ee
As we will see later, $N^{(1)}$ and $N^{(2)}$ will produce terms with polarization power one and two respectively.
The $N^{(1)}$ part is proportional to the vector current $J_{\hat \varepsilon_1}$
\begin{align}\label{eq:NBCJvec}
N^{(1)}(123\cdots n)=\Bigg(\prod _{j=2}^{n-1} \varepsilon_ j\Cdot p_{1\cdots j-1}\Bigg)J_{\hat \varepsilon_1}\,.
\end{align}
This expression agrees with the result given in Ref.~\cite{Chiodaroli:2017ngp}. Remarkably, the $N^{(2)}$ part also takes a very compact form,
\begin{equation}
\label{eq:NBCJ}
N^{(2)}(123\cdots n)=\frac{1}{2} \sum_{i=2}^{n-2}\sum_{\substack{\ell,m=i \\ m>\ell}}^{n-1}(-1)^{\ell-i-1}s_{1\cdots i}\Bigg(\,\prod_{j\in \mathsf{S}_{i\ell m}} \varepsilon_ j \Cdot p_{1\cdots j-1}\Bigg)\!\det(\mathbf{P}_{[i,\ell-1]})J_{\hat \varepsilon_1\otimes\varepsilon_\ell\otimes\varepsilon_m}\,,
\end{equation}
where the set $\mathsf{S}_{i\ell m}=\{2\cdots i-1\}\cup\{\ell+1\cdots\hat{m}\cdots n-1\}$.\footnote{The element with a hat means that it is absent in the set.} For $r_1,r_2\in\{1\cdots n-1\}$, the matrix $\mathbf{P}$ is given by 
\begin{align}\label{eq:PMatrix}
  \mathbf{P}_{r_1r_2}=\left\{\begin{array}{ccc}
  \varepsilon_{r_1}\Cdot p_{1\cdots (r_2+1)} &\quad & r_1\leqslant r_2+1\\
  0 &\quad &\text{otherwise}
  \end{array}\right.\,.
\end{align}
Finally, $\mathbf{P}_{[i,\ell-1]}$ is the submatrix of $\mathbf{P}$ whose rows and columns range between $i$ and $\ell-1$. The number of terms in the determinant $\det(\mathbf{P}_{[i,\ell-1]})$ is $2^{\ell-i-1}$.

The $N^{(1)}$ part given eq.~\eqref{eq:NBCJvec} is generated purely by the first term of eq.~\eqref{eq:vecRule}, and gives all the terms of polarization power one in the BCJ numerator. The $N^{(2)}$ part eq.~\eqref{eq:NBCJ} is a result of all the four fusion rules, and gives all the terms of polarization power two. While the expression of $N^{(1)}$ is rather simple, we now give the explicit formula for $N^{(2)}$ up to seven points as our examples.

 At $n=4$, the summation in eq.~\eqref{eq:NBCJ} has only one term $i=\ell=2$ and $m=3$. The matrix $\mathbf{P}_{[i,\ell-1]}$ is zero-dimensional and in this case we set the determinant to be one. This gives 
\begin{align}\label{eq:NBCJT4}
N^{(2)}(1234)=
-{1\over 2} s_{12}\,J_{\hat\varepsilon_1\otimes\varepsilon _{2}\otimes \varepsilon _{3}}\,.
\end{align}
This result agrees with the Feynman rule calculation in eq.~\eqref{eq:4pqcdbasis123}. The tensor part is in the null space of the four-point propagator matrix as shown previously in eq.~\eqref{eq:4pmab}.

At $n=5$, the nested summation in eq.~\eqref{eq:NBCJ} contains the following four terms: 
\begin{align}\label{eq:fiveBCJT}
N^{(2)}(12345)&=-\frac{1}{2}s_{12}\left(\varepsilon_4\Cdot p_{123}\, J_{\hat \varepsilon_1\otimes\varepsilon_2\otimes\varepsilon_3}+\varepsilon_3\Cdot p_{12}\,J_{\hat \varepsilon_1\otimes\varepsilon_2\otimes\varepsilon_4}-\det(\mathbf{P}_{[2,2]})J_{\hat \varepsilon_1\otimes\varepsilon_3\otimes\varepsilon_4}\right)\nonumber\\
&\quad-{1\over 2}s_{123}\, \varepsilon_2\Cdot p_{1}\,J_{\hat \varepsilon_1\otimes\varepsilon_3\otimes\varepsilon_4}\,,
\end{align}
where the determinant is simply $\det(\mathbf{P}_{[2,2]})=\varepsilon_2\Cdot p_{123}$. 

At $n=6$, from eq.~\eqref{eq:NBCJ} we get
\begin{align}
\label{eq:sixBCJNumTen}
& N^{(2)}(123456)=\nn\\
&-\frac{1}{2}\,s_{12}\Big(\varepsilon_4\Cdot p_{123}\,\varepsilon_5\Cdot p_{1234}\,J_{\hat \varepsilon_1\otimes\varepsilon_2\otimes\varepsilon_3}+\,\varepsilon_3\Cdot p_{12} \,\varepsilon_5\Cdot p_{1234}\,J_{\hat \varepsilon_1\otimes\varepsilon_2\otimes\varepsilon_4}+\varepsilon_3\Cdot p_{12}\, \varepsilon_4\Cdot p_{123}J_{\hat \varepsilon_1\otimes\varepsilon_2\otimes\varepsilon_5}\Big) \nn\\
&+\frac{1}{2}\,s_{12}\Big(\det(\mathbf{P}_{[2,2]})(\varepsilon_5\Cdot p_{1234}\,J_{\hat \varepsilon_1\otimes\varepsilon_3\otimes\varepsilon_4}+\varepsilon_4\Cdot p_{123}\,J_{\hat \varepsilon_1\otimes\varepsilon_3\otimes\varepsilon_5}) -\det(\mathbf{P}_{[2,3]})J_{\hat \varepsilon_1\otimes\varepsilon_4\otimes\varepsilon_5}\Big)\nonumber\\
&-\frac{1}{2}\,s_{123} \Big( \varepsilon_2\Cdot p_{1}\,\varepsilon_5\Cdot p_{1234}\,J_{\hat \varepsilon_1\otimes\varepsilon_3\otimes\varepsilon_4}+\varepsilon_2\Cdot p_{1}\,\varepsilon_4\Cdot p_{123}\,J_{\hat \varepsilon_1\otimes\varepsilon_3\otimes\varepsilon_5}-\varepsilon_2\Cdot p_{1}\,\det(\mathbf{P}_{[3,3]})J_{\hat \varepsilon_1\otimes\varepsilon_4\otimes\varepsilon_5}\Big)\nn\\
&  -{1\over 2}\,s_{1234}\, \varepsilon_2\Cdot p_{1}\,\varepsilon_3\Cdot p_{12}\,J_{\hat \varepsilon_1\otimes\varepsilon_4\otimes\varepsilon_5}\,,
\end{align}
in which two new matrices appear, 
\begin{align}
  \mathbf{P}_{[3,3]}=\,[\varepsilon_3\Cdot p_{1234}]\,,&&
\mathbf{P}_{[2,3]}=\begin{bmatrix}
  \,\varepsilon_2\Cdot p_{123}\, & \,\varepsilon_2\Cdot p_{1234}\, \\
  \,\varepsilon_3\Cdot p_{123}\, & \,\varepsilon_3\Cdot p_{1234}\,
  \end{bmatrix}\,.\nn
\end{align}
As our last example, at $n=7$, eq.~\eqref{eq:NBCJ} gives
\begin{align}
&N^{(2)}(1234567)=\nn\\
&-\frac{1}{2}\,s_{12}\,\Big(\varepsilon_4\Cdot p_{123}\, \varepsilon_5\Cdot p_{1234}\, \varepsilon_6\Cdot p_{12345}\,J_{\hat\varepsilon_1\otimes \varepsilon _2\otimes \varepsilon _3} + \varepsilon_3\Cdot p_{12} \,\varepsilon_5\Cdot p_{1234}\, \varepsilon_6\Cdot p_{12345}\,J_{\hat\varepsilon_1\otimes \varepsilon _2\otimes \varepsilon _4}\Big)\nonumber\\
&-\frac{1}{2}\,s_{12}\,\Big(  \varepsilon_3\Cdot p_{12} \,\varepsilon_4\Cdot p_{123}\,\varepsilon_6\Cdot p_{12345} \,J_{\hat\varepsilon_1\otimes \varepsilon _2\otimes \varepsilon _5} +\varepsilon_3\Cdot p_{12} \,\varepsilon_4\Cdot p_{123}\, \varepsilon_5\Cdot p_{1234} \, J_{\hat\varepsilon_1\otimes \varepsilon _2\otimes \varepsilon _6}\Big) \nonumber\\
&+ \frac{1}{2}\,s_{12}\,\det(\mathbf{P}_{[2,2]})\Big(\varepsilon_5\Cdot p_{1234}\, \varepsilon_6\Cdot p_{12345}  J_{\hat\varepsilon_1\otimes \varepsilon _3\otimes \varepsilon _4} +  \varepsilon_4\Cdot p_{123} \varepsilon_6\Cdot p_{12345} \, J_{\hat\varepsilon_1\otimes \varepsilon _3\otimes \varepsilon _5}\Big) \nonumber\\
&+ \frac{1}{2}\,s_{12}\,\Big( \det(\mathbf{P}_{[2,2]})\, \varepsilon_4\Cdot p_{123} \,\varepsilon_5\Cdot p_{1234} \, J_{\hat\varepsilon_1\otimes \varepsilon _3\otimes \varepsilon _6}- \det(\mathbf{P}_{[2,3]}) \, \varepsilon_6\Cdot p_{12345} \, J_{\hat\varepsilon_1\otimes \varepsilon _4\otimes \varepsilon _5}\Big)\nn \\
&- \frac{1}{2}\,s_{12}\,\Big( \, \det(\mathbf{P}_{[2,3]})\, \varepsilon_5\Cdot p_{1234} \, J_{\hat\varepsilon_1\otimes \varepsilon _4\otimes \varepsilon _6} - \, \det(\mathbf{P}_{[2,4]}) \,   J_{\hat\varepsilon_1\otimes \varepsilon _5\otimes \varepsilon _6}\Big) \nonumber\\
& -{1 \over 2}\,s_{123}\,\varepsilon_2\Cdot p_{1}\,   \Big(\varepsilon_5\Cdot p_{1234} \,  \varepsilon_6\Cdot p_{12345} \, J_{\hat\varepsilon_1\otimes \varepsilon _3\otimes \varepsilon _4} +   \varepsilon_4\Cdot p_{123}\, \varepsilon_6\Cdot p_{12345}  \,J_{\hat\varepsilon_1\otimes \varepsilon _3\otimes \varepsilon _5} \Big) \nonumber\\
&-{1\over 2}\,s_{123} \, \varepsilon_2\Cdot p_{1} \,\Big( \varepsilon_4\Cdot p_{123} \,\varepsilon_5\Cdot p_{1234} \, J_{\hat\varepsilon_1\otimes \varepsilon _3\otimes \varepsilon _6}- \, \det(\mathbf{P}_{[3,3]})\, \varepsilon_6\Cdot p_{12345} \, J_{\hat\varepsilon_1\otimes \varepsilon _4\otimes \varepsilon _5}\Big) \nn\\
&+{1 \over 2}\,s_{123} \, \varepsilon_2\Cdot p_{1}\, \Big( \det(\mathbf{P}_{[3,3]})\, \varepsilon_5\Cdot p_{1234} \, J_{\hat\varepsilon_1\otimes \varepsilon _4\otimes \varepsilon _6} -\det(\mathbf{P}_{[3,4]}) \, J_{\hat\varepsilon_1\otimes \varepsilon _5\otimes \varepsilon _6} \Big) \nonumber\\
&-{1\over 2}\,s_{1234}\, \varepsilon_2\Cdot p_{1} \, \varepsilon_3\Cdot p_{12} \,\Big( \varepsilon_6\Cdot p_{12345} \, J_{\hat\varepsilon_1\otimes \varepsilon _4\otimes \varepsilon _5} + \varepsilon_5\Cdot p_{1234} \, J_{\hat\varepsilon_1\otimes \varepsilon _4\otimes \varepsilon _6}-  \det(\mathbf{P}_{[4,4]}) \, J_{\hat\varepsilon_1\otimes \varepsilon _5\otimes \varepsilon _6} \Big)\nonumber\\
&-{1\over 2}\,s_{12345}\, \varepsilon_2\Cdot p_{1} \, \varepsilon_3\Cdot p_{12} \, \varepsilon_4\Cdot p_{123} \, J_{\hat\varepsilon_1\otimes \varepsilon _5\otimes \varepsilon _6}\,,
\end{align}
where the three new matrices are given by
\begin{align}
\begin{split}
\mathbf{P}_{[4,4]} &=\,[\varepsilon_4\Cdot p_{12345}]\,, \\ 
\mathbf{P}_{[3,4]} &=\begin{bmatrix}
\varepsilon_3\Cdot p_{1234}\, & \,\varepsilon_3\Cdot p_{12345} \\
\varepsilon_4\Cdot p_{1234}\, & \,\varepsilon_4\Cdot p_{12345}
\end{bmatrix}\,,
\end{split}
\begin{split}
\mathbf{P}_{[2,4]}=\begin{bmatrix}
\varepsilon_2\Cdot p_{123}\, & \,\varepsilon_2\Cdot p_{1234}\, & \,\varepsilon_2\Cdot p_{12345} \\
\varepsilon_3\Cdot p_{123}\, & \,\varepsilon_3\Cdot p_{1234}\, & \,\varepsilon_3\Cdot p_{12345} \\
0\, &  \,\varepsilon_4\Cdot p_{1234}\, & \,\varepsilon_4\Cdot p_{12345}
\end{bmatrix}.
\end{split}\nn
\end{align}
Note that the BCJ numerators given above still contain the vector and tensor currents. For conventional on-shell expressions, the tensor currents need to be expanded in a basis, followed by the replacement of the vectors with physical polarizations~\eqref{eq:vecreplace}. As we will see in the next section, this process exposes the generalized gauge freedom encoded in the construction. 

\section{Generalized gauge freedom from tensor currents} \label{sec:FreedomBCJ}
BCJ numerators are not unique due to the generalized gauge freedom. 
In this section, we discuss a particular class of such freedom captured by the tensor currents.

We organize the BCJ numerators computed from eq.~\eqref{eq:current2num} according to polarization powers. The polarization-power-one terms given in eq.~\eqref{eq:NBCJvec} are unique and unaffected by the generalized gauge transformations under concern.
The polarization-power-two terms depend on the choice of the irreducible tensor current basis $\mathsf{b}$. Different $\mathsf b$'s lead to different but gauge-equivalent $N^{\mathsf V}_{\mathsf b}$'s that give the same tree level amplitude.

For a given basis $\mathsf{b}$, the tensor current $N^{\mathsf T}_{\mathsf b}$ must be in the null space of the propagator matrix. Given any monomial of the form $\varepsilon_{i_1}\Cdot p_{j_1} \cdots \varepsilon_{i_{n-4}}\Cdot p_{j_{n-4}} J_{\hat \varepsilon_1\otimes \varepsilon_{i_{n-3}}\otimes \varepsilon_{i_{n-2}}}$, its coefficient must vanish independently under the action of the propagator matrix. We call a monomial of such form a \emph{tensor monomial}. A maximal set of the tensor monomials that are linearly independent up to vector currents forms an \emph{irreducible tensor monomial basis}. It is a special case of the irreducible tensor current basis mentioned in section~\ref{sec:framework}, which includes also the $(\varepsilon_i\Cdot p_j)$ prefactors.
From now on, we will work in the irreducible tensor monomial basis without further specification. The label $\mathsf{b}$ will be used to specify different choices of such bases.

The freedom in the choice of the tensor monomial basis originates from the Clifford algebra in eq.~\eqref{eq:CliffordTensor} and accounts for a particular subset of the generalized gauge freedom of the BCJ numerator.

\subsection{Freedom in non-crossing-symmetric numerators}\label{sec:AmbNoCS}
Irreducible tensor monomial basis provides a quantitative characterization of generalized gauge freedom.
To demonstrate this, we first define the following quantity
\begin{align}
	\mathcal{J}_{i,j}(x)\equiv (1-x)J_{\hat \varepsilon_1\otimes \varepsilon_{i}\otimes \varepsilon_{j}}-x J_{\hat \varepsilon_1\otimes \varepsilon_{j}\otimes \varepsilon_{i}}, \quad (i<j)\,.
\end{align}
For $i>j$, we have $\mathcal{J}_{i,j}(x)=-\mathcal{J}_{j,i}(1-x)$. The freedom captured by the tensor current $J_{\hat \varepsilon_1\otimes \varepsilon_{i}\otimes \varepsilon_{j}}$ is then described by the free parameter $x$. 
Let $\{\mathsf{m}_1,\mathsf{m}_2,\ldots,\mathsf{m}_{H_n}\}$ denote a basis of tensor monomials in the ascending order.
We can lift all the $J_{\hat \varepsilon_1\otimes \varepsilon_{i}\otimes \varepsilon_{j}}$ into $\mathcal{J}_{i,j}(x_h)$ to form a new basis
\begin{align}\label{eq:GenBasis}
  \big\{\mathsf m_1(x_1)\,,~\mathsf m_2(x_2)\,, \ldots, \, \mathsf m_{H_n}(x_{H_n})\big\}
\end{align}
characterized by the vector $[x_1,x_2,\ldots, x_{H_n}]$, where $H_n$ is the dimension of the basis for $n$ external particles. In particular, when all the $x_h$'s are zero, we can write $N(\sigma)$ as
\begin{align}\label{eq:NBCJh0}
	N(\sigma)=N^{(1)}(\sigma)+N^{(2)}(\sigma)=\sum_{h=1}^{H_n}R_{h}(\sigma)\,\mathsf{m}_h(0)+N^{\mathsf V}_{[0,0,\ldots,0]}(\sigma)\,,
\end{align}
where $R_h(\sigma)$ denotes the coefficient of $\mathsf m_h(0)$. 
The subscript of $N^{\mathsf V}_{[0,0,\ldots,0]}$ indicates that it results from the basis~\eqref{eq:GenBasis} with all $x_h$'s being zero. For generic $x_h$'s, we have
\begin{align}\label{eq:GenTV}
	N(\sigma)=N^{\mathsf T}_{[x_1,x_2,\ldots,x_{H_n}]}(\sigma)+N^{\mathsf V}_{[x_1,x_2,\ldots,x_{H_n}]}(\sigma)\,,
\end{align}
where the tensor and vector currents are given by
\begin{align}\label{eq:GenTVexplicit}
N^{\mathsf T}_{[x_1,x_2,\ldots,x_{H_n}]}(\sigma)&=\sum_{h=1}^{H_n}R_{h}(\sigma)\,\mathsf{m}_h(x_h)\,, \cr
N^{\mathsf V}_{[x_1,x_2,\ldots,x_{H_n}]}(\sigma)&=\sum_{h=1}^{H_n}x_h R_{h}(\sigma)\left(\mathsf{m}_h(x_h)\Big|_{\mathcal J_{i,j}(x_h)\rightarrow 2\varepsilon_i\cdot\varepsilon_jJ_{\hat \varepsilon_1}}\right)+N^{\mathsf V}_{[0,0,\ldots,0]}(\sigma)\,.
\end{align}
The replacement $\mathcal J_{i,j}(x_h)\rightarrow 2\varepsilon_i\Cdot\varepsilon_jJ_{\hat \varepsilon_1}$ compensates the mismatch in the vector currents from lifting $J_{\hat \varepsilon_1\otimes \varepsilon_i \otimes \varepsilon_j}$ to $\mathcal J_{i,j}(x_h)$.

In eq.~\eqref{eq:GenTVexplicit}, the tensor currents are expanded in terms of the basis element $\mathsf m_h(x_h)$. As we will show in section~\ref{sec:TopInv}, each coefficient $R_h(\sigma)$ satisfies the null-space condition~eq.~\eqref{eq:CondTen} and corresponds to certain BCJ relations. Thus both $N^{\mathsf T}_{[x_1,x_2,\ldots,x_{H_n}]}(\sigma)$ and the $x_h$ dependent part of the BCJ numerator $N^{\mathsf V}_{[x_1,x_2,\ldots,x_{H_n}]}(\sigma)$ satisfy the null space condition.
Following eq.~\eqref{eq:vecreplace}, we can now remove the tensor currents, and $N^{\mathsf V}_{[x_1,x_2,\ldots,x_{H_n}]}(\sigma)$ can be converted to the conventional form of the BCJ numerator by $J_{\hat \varepsilon_1}\rightarrow\hat \varepsilon_1\Cdot\hat \varepsilon_n$. Interestingly, $N^{\mathsf V}_{[x_1,x_2,\ldots,x_{H_n}]}(\sigma)$ contains a set of parameters that span the subset of the generalized gauge freedom related to a Clifford algebra. The number of free parameters is simply $H_n$, the dimension of irreducible tensor monomial basis.

Next, we illustrate with examples the characterization of the generalized gauge freedom using the tensor currents.
\paragraph{Four-point} The irreducible tensor monomial basis is simply $\{\mathcal J_{2,3}(x)\}$, which is one-dimensional.
For $x=0$, the DDM numerators are
\begin{align}
\label{eq:fourPointTen}
\begin{bmatrix}
\,N(1234)\, \\ \,N(1324)\,
\end{bmatrix}=
\begin{bmatrix}
\varepsilon_2\Cdot p_1 \,\varepsilon_3\Cdot p_{12} \,J_{\hat \varepsilon_1} \\
\varepsilon_3\Cdot p_1 \,\varepsilon_2\Cdot p_{13} \,J_{\hat \varepsilon_1}
\end{bmatrix} +
{1\over 2}\begin{bmatrix}  -s_{12}  \\
\,s_{13} 
\end{bmatrix} \mathcal J_{2,3}(0) +
\begin{bmatrix}  0 \\
-s_{13}\, \varepsilon_2\Cdot\varepsilon_3\, J_{\varepsilon _1}\,
\end{bmatrix}\,.
\end{align}
For a generic basis parameterized by $x$, we have
\begin{align}
\begin{bmatrix}
\,N(1234)\, \\ \,N(1324)\,
\end{bmatrix}=
\begin{bmatrix}
\,N^{\mathsf{V}}_x(1234)\, \\ \,N^{\mathsf{V}}_x(1324)\,
\end{bmatrix}+\begin{bmatrix}
\,N^{\mathsf{T}}_x(1234)\, \\ \,N^{\mathsf{T}}_x(1324)\,
\end{bmatrix}\,,
\end{align}
where
\begin{align}
\begin{bmatrix}
\,N^{\mathsf{T}}_x(1234)\, \\ \,N^{\mathsf{T}}_x(1324)\,
\end{bmatrix}&=
{1\over 2}\begin{bmatrix}  -s_{12}  \\
\,s_{13} 
\end{bmatrix} \mathcal J_{2,3}(x), \cr
\begin{bmatrix}
\,N^{\mathsf{V}}_x(1234)\, \\ \,N^{\mathsf{V}}_x(1324)\,
\end{bmatrix}&=\begin{bmatrix}
\,\varepsilon_2\Cdot p_1 \,\varepsilon_3\Cdot p_{12} \,J_{\hat \varepsilon_1} \\
\,\varepsilon_3\Cdot p_1 \,\varepsilon_2\Cdot p_{13} \,J_{\hat \varepsilon_1}
\end{bmatrix} +
x
\begin{bmatrix}  -s_{12}  \\
\,s_{13} 
\end{bmatrix} \varepsilon_2\Cdot\varepsilon_3\, J_{\varepsilon _1}
+
\begin{bmatrix}  0 \\
-s_{13}\, \varepsilon_2\Cdot\varepsilon_3\, J_{\varepsilon _1}\,
\end{bmatrix}\,.
\end{align}
Clearly, both $N^{\mathsf{T}}_x$ and the $x$-dependent part in $N^{\mathsf{V}}_x$ are in the null space of the four-point propagator matrix, for any value of $x$. The generalized gauge freedom contained in $N^{\mathsf{V}}_x$ is one-dimensional, characterized by the parameter $x$. We can obtain conventional BCJ numerators by the replacement $J_{\hat \varepsilon_1}\rightarrow\hat \varepsilon_1\Cdot\hat\varepsilon_4$ in $N^{\mathsf{V}}_x$.

\paragraph{Five-point} From eq.~\eqref{eq:fiveBCJT}, one can find nine irreducible tensor monomials.
Lifting every $J_{\hat \varepsilon_1\otimes \varepsilon_i \otimes \varepsilon_j}$ to the respective $\mathcal J_{i,j}(x_h)$ leads to one possible choice for the irreducible tensor monomial basis that encodes the information of the generalized gauge freedom. This tensor monomial basis is spanned by
\begin{gather}\label{eq:fiveTMBasis0}
  \varepsilon_4\Cdot p_{1}\, \mathcal J_{2,3}(x_1)\,,~~~ 
  \varepsilon_3\Cdot p_{1}\, \mathcal J_{2,4}(x_2)\,, ~~~
  \varepsilon_2\Cdot p_{1}\, \mathcal J_{3,4}(x_3)\,, ~~~ \varepsilon_4\Cdot p_{2}\, \mathcal J_{2,3}(x_4)\,, ~~~ \varepsilon_4\Cdot p_{3}\, \mathcal J_{2,3}(x_5)\,, ~~~ \nonumber\\
  \varepsilon_3\Cdot p_{2}\, \mathcal J_{2,4}(x_6)\,,~~~\varepsilon _3 \Cdot p_{4}\,  \mathcal J_{2,4}(x_7)\,, ~~~
  \varepsilon_2\Cdot p_{3}\, \mathcal J_{3,4}(x_8)\,, ~~~
   \varepsilon _2 \Cdot p_{4}\, \mathcal J_{3,4}(x_9)\,.
\end{gather}
In this basis, the BCJ numerators given by eq.~\eqref{eq:GenTVexplicit} are
\begin{align}\label{eq:fiveTensorNum}
N^{\mathsf{T}}_{[x_1,\ldots, x_9]}(\sigma)&=\frac{1}{2}\big[R_1 (\sigma)\varepsilon_4\Cdot p_{1}\, \mathcal J_{2,3}(x_1)+R_2 (\sigma)\varepsilon_3\Cdot p_{1}\, \mathcal J_{2,4}(x_2)+R_3 (\sigma)\varepsilon_2\Cdot p_{1}\, \mathcal J_{3,4}(x_3)\nonumber\\
&\quad+R_4 (\sigma)\varepsilon_4\Cdot p_{2}\, \mathcal J_{2,3}(x_4)+R_5 (\sigma)\varepsilon_4\Cdot p_{3}\, \mathcal J_{2,3}(x_5)+R_6 (\sigma)\varepsilon_3\Cdot p_{2}\, \mathcal J_{2,4}(x_6)\nonumber\\
&\quad+R_7 (\sigma) \varepsilon _3 \Cdot p_{4}  \mathcal J_{2,4}(x_7)+R_8 (\sigma)\varepsilon_2\Cdot p_{3}\, \mathcal J_{3,4}(x_8)+R_9 (\sigma) \varepsilon _2 \Cdot p_{4}\, \mathcal J_{3,4}(x_9)\big]\,,\nonumber\\
N^{\mathsf{V}}_{[x_1,\ldots, x_9]}(\sigma)&=\left[x_1R_1 (\sigma)\varepsilon_4\Cdot p_{1}\,+x_4 R_4 (\sigma)\varepsilon_4\Cdot p_{2}+x_5 R_5 (\sigma)\varepsilon_4\Cdot p_{3}\right] \varepsilon_2\Cdot\varepsilon_3\, J_{\hat \varepsilon_1}\nonumber\\
&\quad+ \left[ x_2 R_2 (\sigma)\varepsilon_3\Cdot p_{1}+x_6R_6 (\sigma)\varepsilon_3\Cdot p_{2}+x_7R_7 (\sigma) \varepsilon _3 \Cdot p_{4} \right]\varepsilon_2\Cdot\varepsilon_4\, J_{\hat \varepsilon_1}\cr
&\quad+  \left[ x_3 R_3 (\sigma)\varepsilon_2\Cdot p_{1} +x_8 R_8(\sigma)\varepsilon_2\Cdot p_{3}+ x_9 R_9 (\sigma) \varepsilon _2 \Cdot p_{4} \right] \varepsilon_3\Cdot\varepsilon_4\, J_{\hat \varepsilon_1}\nonumber\\
&\quad+N^{\mathsf{V}}_{[0,\ldots, 0]}(\sigma) \,.
\end{align}
The $N^{\mathsf V}_{[0,0,\ldots, 0]}(\sigma)$ in the above equation takes the form 
\begin{align}
N^{\mathsf{V}}_{[0,\ldots, 0]}=N^{(1)}+
\!\begin{bmatrix}
0 \\ 
s_{12} \,\varepsilon _2\Cdot p_{14}   \,\varepsilon _3\Cdot \varepsilon _4 - s_{124}  \,\varepsilon_2\Cdot p_{1} \,\varepsilon _3\Cdot \varepsilon _4 \\
-s_{13}  \,\varepsilon_4\Cdot p_{123} \, \varepsilon _2\Cdot \varepsilon _3 \\
-s_{13}  \, \varepsilon _4 \Cdot p_{13} \, \varepsilon _2\Cdot \varepsilon _3  +s_{13}  \, \varepsilon _3 \Cdot p_{14} \, \varepsilon _2\Cdot \varepsilon _4 -s_{134} \, \varepsilon _3 \Cdot p_1  \varepsilon _2\Cdot \varepsilon _4 \\
- s_{14} \, \varepsilon _3\Cdot p_{124} \, \varepsilon _2\Cdot \varepsilon _4 - s_{14} \, \varepsilon _2\Cdot  p_{14} \, \varepsilon _3\Cdot \varepsilon _4 \\
\,(s_{14} \, \varepsilon _4 \Cdot p_{13}-s_{134}\,\varepsilon_4\Cdot p_1) \, \varepsilon _2\Cdot \varepsilon _3  - s_{14}\,(\varepsilon _3 \Cdot p_{14}  \, \varepsilon _2\Cdot \varepsilon _4  + \varepsilon _2 \Cdot p_{134}\, \varepsilon _3\Cdot \varepsilon _4)  \,
\end{bmatrix}\!J_{\hat \varepsilon_1}\,,\nn
\end{align}
where the columns are indexed by $\{12345,12435,13245,13425, 14235, 14325\}$, the color orderings in the five-point DDM basis. The coefficients $R_h$ are
\begin{gather}\label{eq:Rh5pt}
R_1=
\begin{bmatrix}
	-s_{12} \cr
-s_{12}\cr
s_{13}\cr
s_{13}\cr
-s_{124}+s_{14}\cr
s_{134}-s_{14}
\end{bmatrix},~
R_2=
\begin{bmatrix}
	-s_{12} \cr
-s_{12}\cr
s_{13}-s_{123}\cr
s_{134}-s_{13}\cr
s_{14}\cr
s_{14}
\end{bmatrix},~
R_3=
\begin{bmatrix}
s_{12}-s_{123} \cr
-s_{12}-s_{124}\cr
-s_{13}\cr
-s_{13}\cr
s_{14}\cr
s_{14}
\end{bmatrix},~
R_4=
\begin{bmatrix}
	-s_{12} \cr
-s_{12}\cr
s_{13}\cr
0\cr
s_{14}\cr
0
\end{bmatrix},\nonumber \\
R_5=
\begin{bmatrix}
	-s_{12} \cr
0\cr
s_{13}\cr
s_{13}\cr
0\cr
-s_{14}
\end{bmatrix},~
R_6=
\begin{bmatrix}
	-s_{12} \cr
-s_{12}\cr
s_{13}\cr
0\cr
s_{14}\cr
0
\end{bmatrix},~
R_7=
\begin{bmatrix}
0 \cr
-s_{12}\cr
0\cr
-s_{13}\cr
s_{14}\cr
s_{14}
\end{bmatrix},~
R_8=
\begin{bmatrix}
s_{12} \cr
0\cr
-s_{13}\cr
-s_{13}\cr
0\cr
s_{14}
\end{bmatrix},~
R_9=
\begin{bmatrix}
0\cr
-s_{12}\cr
0\cr
-s_{13}\cr
s_{14}\cr
s_{14}
\end{bmatrix}.
\end{gather}
All the $R_h$'s correspond to certain BCJ relations to be studied in Section~\ref{sec:TopInv}. Therefore, the coefficients of each $x_h$ in both $N^{\mathsf T}_{[x_1,\ldots, x_9]}$ and $N^{\mathsf V}_{[x_1,\ldots, x_9]}$ are in the null space of the five-point propagator matrix. We can now remove $N^{\mathsf T}_{[x_1,\ldots, x_9]}$ and translate $N^{\mathsf V}_{[x_1,\ldots, x_9]}$ into conventional BCJ numerators by the replacement $J_{\hat \varepsilon_1} \rightarrow \hat \varepsilon_1\Cdot \hat\varepsilon_5$. The generalized gauge freedom captured by the tensor currents is characterized by the nine free $x_h$'s.

\paragraph{Higher multiplicities}
The pattern shown above can be carried over to higher multiplicities. We can always construct the basis~\eqref{eq:GenBasis} by reading off the irreducible tensor monomials from eq.~\eqref{eq:NBCJ}, and promote each $J_{\hat \varepsilon_1\otimes \varepsilon_i \otimes \varepsilon_j}$ to $\mathcal J_{i,j}(x_h)$.
The parameters $\{ x_1,\cdots, x_{H_n} \} $ control the subset of the generalized gauge freedom that are induced by a Clifford algebra. We show the dimension of such gauge freedom up to eight points in the table below:

\begin{center}
\begin{tabular}{c|ccccc}
&$n=4$& $n=5$&$n=6$&$n=7$&$n=8$ \\ \hline
free parameters &1&9&90&1080&15435
\end{tabular}
\end{center}

\subsection{Tensor monomial diagram and topology}\label{sec:TenGraph}
Tensor currents control part of the generalized gauge freedom induced by the Clifford algebra~\eqref{eq:CliffordTensor}, among which we are most interested in those that leaves the numerators crossing symmetric.
Our goal is to find the irreducible tensor monomial bases that leave $N^{\mathsf V}$ crossing symmetric.
To better understand this subset of bases, we introduce \emph{tensor monomial diagram}, a diagrammatic representation of tensor monomials, constructed by the following rules:
\begin{itemize}
\item Each $\varepsilon_i \Cdot p_j$ factor is denoted by a black node $\bullet$ with label $i$ for $j>1$ and by a white node $\circ$ with label $i$ for $j=1$.
\item The tensor $J_{\hat\varepsilon_1\otimes \varepsilon _i\otimes \varepsilon _j}$ is denoted by a thick black line, of which the left endpoint is labeled by $i$ and the right endpoint by $j$.
\item If $\varepsilon_i\Cdot p_j\,\varepsilon_j\Cdot p_k$ appears, we connect the two nodes $i$ and $j$ by a red line. If 
$\varepsilon_i\Cdot p_j \, J_{\hat \varepsilon_1\otimes\varepsilon_j\otimes \varepsilon_k}$ or $\varepsilon_i\Cdot p_k \, J_{\hat \varepsilon_1\otimes\varepsilon_j\otimes \varepsilon_k}$ appears, we draw a red line between the vertex $i$ and the corresponding endpoint of the thick black line.
\item We organize the diagram such that the white node $\circ$ always appears as the left-most element. 
\end{itemize} 
The diagram for a tensor monomial may be fully connected or a collection of several disconnected pieces. In a disconnected diagram, each connected component contains either the thick black line, denoting the tensor current, or a white node. In particular, we call the piece containing the thick line the \emph{tensor part} of the diagram. Stripping off all the labels from a monomial diagram, we get the \emph{topology} of the diagram. Two topologies are considered same if their tensor parts are identical up to a left-right reflection, and the pieces connected to white nodes are identical. As an example, we show the diagram and topology of a particular monomial:
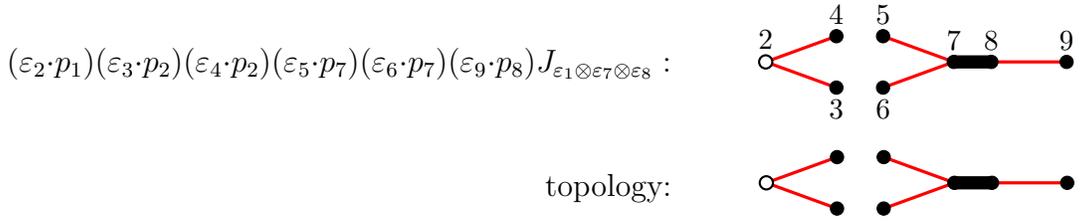
\begin{align*}
(\varepsilon _2\Cdot p_{1})  (\varepsilon _3\Cdot p_{2})   (\varepsilon _4\Cdot p_{2})   (\varepsilon _5\Cdot p_{7})   (\varepsilon _6\Cdot p_{7})  (\varepsilon _9\Cdot p_{8}) J_{\varepsilon _1\otimes \varepsilon _7\otimes \varepsilon _8}:& & & & & \begin{tikzpicture}[baseline={([yshift=-0.7ex]current bounding box.center)}]
\tikzstyle{every node}=[font=\small]
\draw [very thick,red] (0,0) -- (20:1);
\filldraw [black] (20:1) circle (2.5pt) node [above=0.5pt]{$4$};
\draw [very thick,red] (0,0) -- (-20:1);
\filldraw [black] (-20:1) circle (2.5pt) node [below=0.5pt]{$3$};
\filldraw [thick,fill=white,draw=black] (0,0) circle (2.5pt) node [above=0.5pt]{$2$};
\begin{scope}[xshift=2.5cm]
\draw [very thick,red] (-160:1) -- (0,0) -- (160:1);
\draw [line width=5pt] (0,0) -- (0.5,0);
\draw [very thick,red] (1.5,0) -- (0.5,0);
\filldraw [black] (160:1) circle (2.5pt) node [above=0.5pt]{$5$};
\filldraw [black] (-160:1) circle (2.5pt) node [below=0.5pt]{$6$};
\filldraw [black] (1.5,0) circle (2.5pt) node [above=0.5pt]{$9$};
\filldraw [black] (0.5,0) circle (2.5pt) node [above=0.5pt]{$8$};
\filldraw [black] (0,0) circle (2.5pt) node [above=0.5pt]{$7$};
\end{scope}
\end{tikzpicture}\nonumber\\
\text{topology:}& & & & &\begin{tikzpicture}[baseline={([yshift=-0.ex]current bounding box.center)}]
\draw [very thick,red] (0,0) -- (20:1);
\filldraw [black] (20:1) circle (2.5pt);
\draw [very thick,red] (0,0) -- (-20:1) node[below=0.5pt,opacity=0]{$3$};
\filldraw [black] (-20:1) circle (2.5pt);
\filldraw [thick,fill=white,draw=black] (0,0) circle (2.5pt) node [above=0.5pt]{\phantom{$2$}};
\begin{scope}[xshift=2.5cm]
\draw [very thick,red]  (-160:1) -- (0,0) -- (160:1);
\draw [line width=5pt] (0,0) -- (0.5,0);
\draw [very thick,red] (1.5,0) -- (0.5,0);
\filldraw [black] (160:1) circle (2.5pt);
\filldraw [black] (-160:1) circle (2.5pt);
\filldraw [black] (1.5,0) circle (2.5pt) node [above=0.5pt]{\phantom{$9$}};
\filldraw [black] (0,0) circle (2.5pt);
\filldraw [black] (0.5,0) circle (2.5pt);
\end{scope}
\end{tikzpicture}\,.
\end{align*}
The diagrams with labels are in a one-to-one correspondence with the tensor monomials, while each topology characterizes a minimal set of tensor monomials that is mapped to itself under permutations of labels.


\subsection{Freedom for crossing-symmetric numerators}\label{sec:AmbCS}

Since $N(\sigma)$ is crossing symmetric by construction, crossing-symmetric $N^{\mathsf V}_{\mathsf b}$ is the consequence of a crossing-symmetric tensor current $N^{\mathsf T}_{\mathsf b}$ and basis $\mathsf b$.
The symmetry property of tensor diagrams enables us to easily construct such bases 
and study the generalized gauge freedom therein.  
In practice, we start with the basis~\eqref{eq:GenBasis} with all the $x_h=0$ and extract the topologies for these tensor monomials. Then the following two rules select the basis that respect the crossing symmetries, based on whether the tensor part of the given topology is left-right symmetric:
\begin{enumerate}[label=(\arabic*)]
\item if the tensor part is symmetric under the left-right reflection, we replace $J_{\hat\varepsilon_1\otimes \varepsilon_i\otimes \varepsilon _j}$ by $\mathcal{J}_{i,j}({1/ 2})$ in each tensor monomial of this topology.
\item if the tensor part is \emph{not} symmetric under the left-right reflection, we divide the diagrams in this topology into two sets such that the tensor parts of one set are mapped into the other under the left-right reflection. Then we replace $J_{\hat \varepsilon_1\otimes\varepsilon_i\otimes\varepsilon_j}$ by $\mathcal{J}_{i,j}(x)$ in one set and by $\mathcal{J}_{i,j}(1-x)$ in the other.
\end{enumerate}
Imposing crossing symmetries reduces the number of free parameters from $H_n$ to half of the number of the topologies with left-right asymmetric tensor part. This construction covers a subspace of crossing-symmetric BCJ numerators.
We illustrate the construction with several examples:


\paragraph{Four-point} There is only one independent tensor monomial and its topology is \tikz{\pic{box};}, which enjoys the left-right reflection symmetry. We thus choose $\{ \mathcal J_{2, 3}(\frac{1}{2}) \}$ as the basis, and write the numerator as
\begin{align}\label{eq:fourBCJNumT2V}
\left[\begin{matrix}
N(1234)\cr N(1324)
\end{matrix}\right]
=
\begin{bmatrix}
\varepsilon_2\Cdot p_1 \varepsilon_3\Cdot p_{12} J_{\hat \varepsilon_1} \\
\varepsilon_3\Cdot p_1 \varepsilon_2\Cdot p_{13} J_{\hat \varepsilon_1}
\end{bmatrix}+
{1\over 2}\left[\begin{matrix}
 -s_{12} \mathcal J_{2, 3}(\frac{1}{2})-s_{12}\varepsilon_2\Cdot\varepsilon_3J_{\varepsilon _1}\cr
  s_{13}\mathcal J_{2, 3}(\frac{1}{2})-s_{13}\varepsilon_2\Cdot\varepsilon_3J_{\varepsilon _1}
\end{matrix}\right].
\end{align}
We can remove the tensors, as they are in the null space of the propagator matrix. The numerator then becomes
\begin{align}\label{eq:fourPvec2}
\left[\begin{matrix}
N(1234)\cr N(1324)
\end{matrix}\right]\xrightarrow{\texttt{ remove tensors }}
\left[\begin{matrix}
N^{\mathsf V}_{1\over 2}(1234)\cr N^{\mathsf V}_{1\over 2}(1324)
\end{matrix}\right]=
\begin{bmatrix}
\varepsilon_2\Cdot p_1 \varepsilon_3\Cdot p_{12} J_{\hat \varepsilon_1} \\
\varepsilon_3\Cdot p_1 \varepsilon_2\Cdot p_{13} J_{\hat \varepsilon_1}
\end{bmatrix}+{1\over 2}
\left[\begin{matrix}
 -s_{12} \varepsilon_2\Cdot\varepsilon_3J_{\varepsilon _1}\cr
 -s_{13}  \varepsilon_2\Cdot\varepsilon_3J_{\varepsilon _1}
\end{matrix}\right].
\end{align}
Clearly the result is symmetric under the relabeling $2\leftrightarrow 3$. In this case, there is no gauge freedom induced by the Clifford algebra.

\paragraph{Five-point} In this case, there are two distinct topologies, 
\begin{subequations}
\begin{align}
\label{eq:fivePa}
&  \begin{tikzpicture}[baseline={([yshift=-0.8ex]current bounding box.center)}]
	\filldraw [thick,fill=white,draw=black] (0,0) circle (2.5pt);
	\draw [line width=5pt] (1,0) -- (1.5,0);
	\filldraw [fill=black] (1,0) circle (2.5pt);
	\filldraw [fill=black] (1.5,0) circle (2.5pt);
\end{tikzpicture} & &\text{number of monomials: }\frac{3!}{2}=3
\,, \\
\label{eq:fivePb}
& \begin{tikzpicture}[baseline={([yshift=-0.8ex]current bounding box.center)}]
\draw [very thick,red] (0,0) -- (1,0);
\draw [line width=5pt] (1,0) -- (1.5,0);
\filldraw [fill=black] (0,0) circle (2.5pt);
\filldraw [fill=black] (1,0) circle (2.5pt);
\filldraw [fill=black] (1.5,0) circle (2.5pt);
\end{tikzpicture} & &\text{number of monomials: }3!=6
\,.
\end{align}
\end{subequations}
The tensor monomials of the topology~\eqref{eq:fivePa} are $\varepsilon_4\Cdot p_1 \,J_{\hat \varepsilon_1\otimes \varepsilon_2 \otimes \varepsilon_3}$, $\varepsilon_3\Cdot p_1 \,J_{\hat \varepsilon_1\otimes \varepsilon_2 \otimes \varepsilon_4}$ and $\varepsilon_2\Cdot p_1 \,J_{\hat \varepsilon_1\otimes \varepsilon_3 \otimes \varepsilon_4}$, all of which have symmetric tensor parts. Hence, the basis for topology~\eqref{eq:fivePa} is
\begin{align}
\label{eq:fivePbasisc}
\big\{\,\varepsilon_4\Cdot p_1\, \mathcal J_{2,3}(\tfrac{1}{2})\,,\quad \varepsilon_3\Cdot p_1 \,\mathcal J_{2,4}(\tfrac{1}{2})\,,\quad\varepsilon_2\Cdot p_1\,\mathcal J_{3,4}(\tfrac{1}{2})\,\big\}\,.
\end{align}
The topology~\eqref{eq:fivePb} contains six tensor monomials, among which the diagrams for $\varepsilon_4\Cdot p_{2} \,J_{\hat \varepsilon_1 \otimes \varepsilon_{2} \otimes \varepsilon_{3}}$, $\varepsilon_3\Cdot p_{2} \,J_{\hat \varepsilon_1 \otimes \varepsilon_{2} \otimes \varepsilon_{4}}$ and $\varepsilon_2\Cdot p_{3} \,J_{\hat \varepsilon_1 \otimes \varepsilon_{3} \otimes \varepsilon_{4}}$ are the mirror images of those for $\varepsilon_4\Cdot p_{3} \,J_{\hat \varepsilon_1 \otimes \varepsilon_{2} \otimes \varepsilon_{3}}$, $\varepsilon_3\Cdot p_{4} \,J_{\hat \varepsilon_1 \otimes \varepsilon_{2} \otimes \varepsilon_{4}}$ and $\varepsilon_2\Cdot p_{4} \,J_{\hat \varepsilon_1 \otimes \varepsilon_{3} \otimes \varepsilon_{4}}$. Since the tensor part is asymmetric, we choose the basis as:
\begin{align}
\label{eq:fivePbasisb}
\big\{\, &\varepsilon_4\Cdot p_{2}\,\mathcal J_{2,3}(x)\,,& &\varepsilon_4\Cdot p_{3}\,\mathcal J_{2,3}(1-x)\,,& &\varepsilon_3\Cdot p_{2}\,  \mathcal J_{2,4}(x)\,,\nonumber\\
&\varepsilon _3\Cdot p_{4}\, \mathcal J_{2,4}(1-x)\,,& &\varepsilon_2\Cdot p_{3}\, \mathcal J_{3,4}(x)\,,& &\varepsilon _2\Cdot p_{4}\, \mathcal J_{3,4}(1-x)\,\big\}.
\end{align}
Eq.~\eqref{eq:fivePbasisc} and~\eqref{eq:fivePbasisb} together form a crossing-symmetric basis, in which the BCJ numerator is also crossing symmetric:
\begin{align}
&N^{\mathsf{V}}_{[\frac{1}{2},\frac{1}{2},{1\over 2}, x, 1-x,y,1-x,x,1-x]}(12345) \\
&\quad=N^{(1)}(12345)-{1\over 2}s_{12} \left[\varepsilon_4\Cdot p_{1}\,+2x \varepsilon_4\Cdot p_{2}+(2-2x) \varepsilon_4\Cdot p_{3}\right]\varepsilon_2\Cdot\varepsilon_3\, J_{\hat \varepsilon_1}\nonumber\\
&\quad\quad-{1\over 2}s_{12}(\varepsilon_3\Cdot p_{1}+2x\varepsilon_3\Cdot p_{2} )\varepsilon_2\Cdot\varepsilon_4\, J_{\hat \varepsilon_1}+{1\over 2}\big[(-s_{123}+s_{12})\varepsilon_2\Cdot p_{1} + 2x s_{12}\varepsilon_2\Cdot p_{3}\big]\varepsilon_3\Cdot\varepsilon_4\, J_{\hat \varepsilon_1}\,,\nonumber
\end{align}
where $N^{(1)}$ is given by eq.~\eqref{eq:NBCJvec}. We only have one free parameter $x$ here.

\paragraph{Six-point} In this case, there are six distinct topologies, as shown in table~\ref{fig:sixTopfine}. 
The three topologies in the left column have left-right symmetric tensor parts, to which we just perform the replacement $J_{\hat\varepsilon_1\otimes \varepsilon _{i}\otimes \varepsilon _{j}}\rightarrow \mathcal J_{i,j}({1\over 2})$. The three topologies in the right column have asymmetric tensor parts. For each such topology $a$, we divide its monomial diagrams into two groups with the tensor parts in one group being the mirror of those in the other. We apply $J_{\hat\varepsilon_1\otimes \varepsilon _{i}\otimes \varepsilon _{j}}\rightarrow \mathcal J_{i,j}(x_a)$ to one group and $J_{\hat\varepsilon_1\otimes \varepsilon _{i}\otimes \varepsilon _{j}}\rightarrow \mathcal J_{i,j}(1-x_a)$ to the other. As a result, the bases for these three topologies are
\begin{align}
&\big\{\varepsilon _{\sigma_2}\Cdot p_{\sigma_3} \, \varepsilon _{\sigma_3}\Cdot p_{\sigma_5}\, \mathcal J_{\sigma_4,\sigma_5}(x_1)\,,\,\varepsilon _{\sigma_2}\Cdot p_{\sigma_3}\, \varepsilon _{\sigma_3}\Cdot p_{\sigma_4}\, \mathcal J_{\sigma_4,\sigma_5}(1-x_1)  \,|\,\sigma_4<\sigma_5\big\}\,,\nonumber\\
&\big\{\varepsilon _{\rho_4}\Cdot p_{\rho_3}\, \varepsilon _{\rho_5}\Cdot p_{\rho_3}\, \mathcal J_{\rho_2,\rho_3}(x_2)\,,\, \varepsilon _{\rho_4}\Cdot p_{\rho_2}\, \varepsilon _{\rho_5}\Cdot p_{\rho_2}\, \mathcal J_{\rho_2,\rho_3}(1-x_2)\,|\,\rho_2<\rho_3\big\}\,,\nonumber\\
&\big\{\varepsilon _{\omega_4}\Cdot p_{\omega_3}\, \varepsilon _{\omega_5}\Cdot p_1\, \mathcal J_{\omega_2,\omega_3}(x_3)\,,\, \varepsilon _{\omega_4}\Cdot p_{\omega_2}\,\varepsilon _{\omega_5}\Cdot p_1\, \mathcal J_{\omega_2,\omega_3}(1-x_3) \,|\, \omega_2<\omega_3\big\}\,,
\end{align}
where $\sigma$, $\rho$ and $\omega$ are permutations of $\{2,3,4,5\}$ that preserve the relative ordering of the tensor part indices. One can easily check that both the basis and numerators are crossing symmetric. We now have three parameters $\{x_1,x_2,x_3\}$ to control the freedom in the crossing-symmetric numerators.

\begin{table}[t]
\centering
\begin{tikzpicture}
\node at (0.25,1.25) {topology};
\node at (3,1.25) [align=center] {number of \\ monomials};
\draw [thin] (-2.1,0.7) -- (11.1,0.7) (4.25,1.9) -- (4.25,-3.6);
\filldraw [thick,fill=white,draw=black] (160:1) circle (2.5pt) (-160:1) circle (2.5pt);
\draw [line width=5pt] (0,0) -- (0.5,0);
\filldraw [black] (0,0) circle (2.5pt) (0.5,0) circle (2.5pt);
\node at (3,0) {$\displaystyle \frac{4!}{2\times 2}=6$};
\node at (-1.75,0) {(1)};
\begin{scope}[yshift=-1.5cm]
\draw [very thick,red] (-1,0) -- (0,0);
\draw [line width=5pt] (1,0) -- (1.5,0);
\filldraw [black] (0,0) circle (2.5pt) (1,0) circle (2.5pt) (1.5,0) circle (2.5pt);
\filldraw [thick,fill=white,draw=black] (-1,0) circle (2.5pt);
\node at (3,0) {$\displaystyle \frac{4!}{2}=12$};
\node at (-1.75,0) {(3)};
\end{scope}
\begin{scope}[yshift=-3cm]
\draw [very thick,red] (-1,0) -- (0,0) (0.5,0) -- (1.5,0);
\draw [line width=5pt] (0,0) -- (0.5,0);
\filldraw [black] (-1,0) circle (2.5pt) (0,0) circle (2.5pt) (0.5,0) circle (2.5pt) (1.5,0) circle (2.5pt);
\node at (3,0) {$\displaystyle \frac{4!}{2}=12$};
\node at (-1.75,0) {(5)};
\end{scope}
\begin{scope}[xshift=6.5cm]
\node at (0.25,1.25) {topology};
\node at (3,1.25) [align=center] {number of \\ monomials};
\draw [very thick,red] (-1,0) -- (1,0);
\draw [line width=5pt] (1,0) -- (1.5,0);
\filldraw [black] (-1,0) circle (2.5pt) (0,0) circle (2.5pt) (1,0) circle (2.5pt) (1.5,0) circle (2.5pt);
\node at (3,0) {$\displaystyle 4!=24$};
\node at (-1.75,0) {(2)};
\end{scope}
\begin{scope}[xshift=6.5cm,yshift=-1.5cm]
\draw [very thick,red] (160:1) -- (0,0) -- (-160:1);
\draw [line width=5pt] (0,0) -- (0.5,0);
\filldraw [black] (160:1) circle (2.5pt) (-160:1) circle (2.5pt) (0,0) circle (2.5pt) (0.5,0) circle (2.5pt);
\node at (3,0) {$\displaystyle \frac{4!}{2}=12$};
\node at (-1.75,0) {(4)};
\end{scope}
\begin{scope}[xshift=6.5cm,yshift=-3cm]
\draw [very thick,red] (1,0) -- (0,0);
\draw [line width=5pt] (1,0) -- (1.5,0);
\filldraw [black] (0,0) circle (2.5pt) (1,0) circle (2.5pt) (1.5,0) circle (2.5pt);
\filldraw [thick,fill=white,draw=black] (-1,0) circle (2.5pt);
\node at (3,0) {$\displaystyle 4!=24$};
\node at (-1.75,0) {(6)};
\end{scope}
\end{tikzpicture}
\caption{The topology and the number of monomials contained at six points. }
\label{fig:sixTopfine}
\end{table}

\paragraph{Higher multiplicities} The same analysis applies to generic cases, and we list below the number of free parameters in the crossing-symmetric numerators up to eight points.
\begin{center}
\begin{tabular}{c|ccccc}
&$n=4$& $n=5$&$n=6$&$n=7$&$n=8$ \\ \hline
free parameters &0&1&3& 9&22
\end{tabular}
\end{center}

\section{BCJ relations from monomial diagrams}\label{sec:TopInv}
In this section, we demonstrate that, given an irreducible tensor monomial basis, the coefficient vector $R_h$ of each basis element $\mathsf{m}_h$ always corresponds to a BCJ relation. Therefore, each $R_h$ must lie in the null space of the propagator matrix at any multiplicity.

We first work out the rules to write down $R_h$ for the monomials with a connected topology in section~\ref{sec:connected}, while the rules for generic disconnected topologies will be given in section~\ref{sec:disconnected}. The building blocks for these two cases are \emph{binary} and \emph{generalized binary} BCJ relations respectively, both of which will be defined shortly.

We will work in the basis with all $x_h$'s being zero in eq.~\eqref{eq:GenBasis}. However, the discussion is completely generic since $R_h$ does not depend on $x_h$, and thus it remains the same in other bases.

\subsection{Binary BCJ relations and connected topologies}\label{sec:connected}

A binary BCJ relation is defined as the following object being zero on shell
\begin{align}\label{def:binaryBCJ}
  \mathcal{B}(\alpha_1\alpha_2\cdots\alpha_{n-2})&\equiv\left(\mathbb{I}-\mathbb{P}_{(\alpha_1\alpha_2\cdots\alpha_{n-2})}\right)\cdot\left(\mathbb{I}-\mathbb{P}_{(\alpha_1\cdots\alpha_{n-3})}\right)\cdots\left(\mathbb{I}-\mathbb{P}_{(\alpha_1\alpha_2)}\right)\nonumber\\
  &\quad\;\cdot\big[s_{1\alpha_1}A(1\alpha_1\cdots\alpha_{n-2}n)\big],
\end{align}
where the operator $\mathbb{P}_{(\alpha_1\alpha_2\cdots\alpha_j)}$ generates the cyclic permutation $(\alpha_1\alpha_2\cdots\alpha_j)$ onto the indices of the object it acts on, and $\mathbb{I}$ is the identity permutation. For example, the following action gives
\begin{align*}
  \left(\mathbb{I}-\mathbb{P}_{(\alpha_1\alpha_2)}\right)\cdot\big[s_{1\alpha_1}A(1,\alpha_1,\alpha_2\cdots)\big]=s_{1\alpha_1}A(1\alpha_1\alpha_2\cdots)-s_{1\alpha_2}A(1\alpha_2\alpha_1\cdots)\,.
\end{align*}
When generating $\mathcal{B}(\alpha_1\alpha_2\cdots\alpha_{n-2})$, we start with a single object $s_{1\alpha_1}A(1\alpha_1\alpha_2\cdots)$, while each action of $(\mathbb{I}-\mathbb{P}_{(\alpha_1\alpha_2\cdots)})$ doubles the number of terms. Therefore, the total number of terms involved in the binary BCJ relation $\mathcal{B}(\alpha_1\alpha_2\cdots\alpha_{n-2})=0$ is $2^{n-3}$. The algebraic construction in eq.~\eqref{def:binaryBCJ} leads to the following closed form
\begin{align}
\mathcal{B}(\alpha_1\alpha_2\cdots\alpha_{n-2})&=\sum_{i=1}^{n-2}\sum_{\rho\in\{\alpha_{i-1},\cdots,\alpha_1\}\shuffle \{\alpha_{i+1},\cdots,\alpha_{n-2}\}} (-1)^{i-1}s_{1\alpha_i}A(1,\alpha_i, \rho, n)\,,
\end{align}
where $\shuffle$ denotes the order-preserving shuffle of two sets.\footnote{With a slight abuse of language, we will simply refer $\mathcal{B}(\alpha_1\cdots\alpha_{n-2})$ as a binary BCJ relation. This nomenclature also applies to the other BCJ relations later given in eq.~\eqref{eq:gb} and~eq.~\eqref{eq:BiLocalBCJ}.}

To show that eq.~\eqref{def:binaryBCJ} is indeed a BCJ relation, we express it as a linear combination of the generalized BCJ relations~\cite{Bern:2008qj,Feng:2010my,BjerrumBohr:2010ta,BjerrumBohr:2010zb,Chen:2011jxa}

\begin{align}\label{eq:gb}
  \mathcal{R}(\beta_1\beta_2\cdots\beta_s | \alpha_1\alpha_2\cdots\alpha_r)\equiv
  \!\!\sum_{\substack{\rho\in\{\beta_1\beta_2\cdots\beta_s\} \\ \quad\quad\shuffle \{\alpha_1\alpha_2\cdots\alpha_r\}}}\sum_{j=1}^r\sum_{\rho_i<\rho_{\alpha_j}}s_{\sigma_i\alpha_j}A(1,\rho,n)\,,
\end{align}
where in last summation we define $\rho_1\equiv 1$. We find that the following identity holds on the support of the momentum conservation
\begin{align}\label{LBCJ2NBCJ}
\mathcal{B}(\alpha_1\cdots\alpha_{n-2})
&=\sum_{i=1}^{n-3}(-1)^{i}\mathcal{R}(\alpha_i\cdots\alpha_2\alpha_1|\alpha_{i+1}\cdots\alpha_{n-2})\,. 
\end{align}
For example, at four points, eq.~\eqref{def:binaryBCJ} gives
\begin{align}
\mathcal{B}(23)=\left(\mathbb{I}-\mathbb{P}_{(23)}\right)\cdot\big[s_{12}A(1234)\big]=s_{12}A(1234)-s_{13}A(1324)\,,
\end{align}
which agrees with the generalized BCJ relation
\begin{align}\label{eq:4pgb}
-\mathcal{R}(2|3)=-(s_{13}+s_{23})A(1234)-s_{13}A(1324)=s_{12}A(1234)-s_{13}A(1324)\,.
\end{align}
Similarly, at five points, we have
\begin{align}\label{eq:five2}
\mathcal{B}(234)&=\left(\mathbb{I}-\mathbb{P}_{(234)}\right)\cdot\left(\mathbb{I}-\mathbb{P}_{(23)}\right)\cdot\left[s_{12}A(12345)\right]\nonumber\\
&=\left(\mathbb{I}-\mathbb{P}_{(234)}\right)\cdot\left[s_{12}A(12345)-s_{13}A(13245)\right]\nonumber\\
&=s_{12}A(12345)-s_{13}A(13245)-s_{13}A(13425)+s_{14}A(14325)\,.
\end{align}
The right hand side of eq.~\eqref{LBCJ2NBCJ} gives the same result
\begin{align}\label{eq:five1}
&\quad -\mathcal{R}(2|34)+\mathcal{R}(32|4)\cr
&=-\left[(s_{1234}-s_{12})A(12345)+(s_{1234}-s_{132}+s_{13})A(13245)+s_{134}A(13425)\right]\nonumber\\
&\quad+(s_{1324}-s_{132})A(13245)+(s_{134}-s_{13})A(13425)+s_{14}A(14325)\nonumber\\
&=s_{12}A(12345)-s_{13}A(13245)-s_{13}A(13425)+s_{14}A(14325)\,.
\end{align}
In both eq.~\eqref{eq:4pgb} and~\eqref{eq:five1}, only momentum conservation has been used. We have checked the validity of eq.~\eqref{LBCJ2NBCJ} up to twenty points.

With the binary BCJ relations in hand, we present the algorithm to read off the coefficients $R_h$'s of monomials with connected topologies. Since these monomials do not contain $\varepsilon_i\Cdot p_1$, their diagrams can be viewed as two trees rooted on the node of the tensor part \tikz{\pic{box};}. We start with the simplest connected topology, the \emph{chain topology}:
\begin{align}
  \begin{tikzpicture}[scale=1.0]
    \draw [red,very thick] (0.5,0) -- (1.5,0) (0,0) -- (-1,0) (2.5,0) -- (3.5,0) (-2,0) -- (-3,0);
    \draw [red,very thick,dashed] (-1,0) -- (-2,0) (1.5,0) -- (2.5,0);
    \filldraw [black] (0,0) circle (2.5pt) node [above=0pt]{$\alpha_{i\vphantom{+}}$} (0.5,0) circle (2.5pt) node [above=0pt]{$~~\alpha_{i+1}$} (1.5,0) circle (2.5pt) node[above=0pt]{$\alpha_{i+2}$} (2.5,0) circle (2.5pt) node[above=0pt]{$\alpha_{n-3}$} (3.5,0) circle (2.5pt) node[above=0pt]{$\alpha_{n-2}$} (-1,0) circle (2.5pt) node[above=0pt]{$\alpha_{i-1}$} (-2,0) circle (2.5pt) node[above=0pt]{$\alpha_2$} (-3,0) circle (2.5pt) node[above=0pt]{$\alpha_1$};
    \draw [line width=5pt] (0,0) -- (0.5,0);
  \end{tikzpicture}\,.
\end{align}
This diagram is associated with the monomial
\begin{align}
  \varepsilon_{\alpha_1}\Cdot p_{\alpha_2} \, \varepsilon_{\alpha_2}\Cdot p_{\alpha_3} \cdots \varepsilon_{\alpha_{i-1}}\Cdot p_{\alpha_i} J_{\hat \varepsilon_1\otimes \varepsilon_{\alpha_i}\otimes \varepsilon_{\alpha_{i+1}}} \, \varepsilon_{\alpha_{i+2}}\Cdot p_{\alpha_{i+1}} \cdots \varepsilon_{\alpha_{n-3}}\Cdot p_{\alpha_{n-4}}\, \varepsilon_{\alpha_{n-2}} \Cdot p_{\alpha_{n-3}}\,.
\end{align}
We find that the coefficient $R_h$ of this monomial appear in a single binary BCJ relation. Namely, we have
\begin{align}\label{eq:chainrule}
  \text{chain topology:}& &\sum_{\sigma\in S_{n-2}}R_h(\sigma)A(\sigma)=\mathcal{B}(\alpha_1\alpha_2\cdots\alpha_{n-2})\,.
\end{align}
As a simple example, we consider the monomial $\varepsilon_2\Cdot p_3 \,J_{\hat \varepsilon_1\otimes\varepsilon_3\otimes\varepsilon_4}$ that appears at five points. Its monomial diagram is
\begin{align}
  \varepsilon_2\Cdot p_3 \,J_{\hat \varepsilon_1\otimes\varepsilon_3\otimes\varepsilon_4}:\qquad\begin{tikzpicture}[baseline={([yshift=-0.8ex]current bounding box.center)}]
  \draw [red,very thick] (-1,0) -- (0,0);
  \filldraw [black] (-1,0) circle (2.5pt) node[above=0pt]{$2$} (0,0) circle (2.5pt) node[above=0pt]{$3$} (0.5,0) circle (2.5pt) node[above=0pt]{$4$};
  \draw [black,line width=5pt] (0,0) -- (0.5,0);
  \end{tikzpicture}\,.
\end{align}
The coefficient vector is $R_8$ in eq.~\eqref{eq:Rh5pt} and it satisfies eq.~\eqref{eq:chainrule}
\begin{align}
  \sum_{\sigma\in S_3}\!R_h(\sigma)A(\sigma)=\mathcal{B}(234)=s_{12}A(12345)-s_{13}A(13245)-s_{13}A(13425)+s_{14}A(14325)\,.
\end{align}

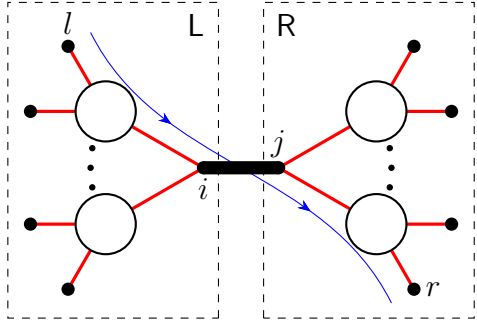
\begin{figure}[t]
	\centering
	\begin{tikzpicture}[ns/.style={circle,fill=black,inner sep=0,minimum size=5pt},arrowmark/.style 2 args={decoration={markings,mark=at position #1 with \arrow{#2}}}]
	\draw [red,very thick] (0,0) node[ns]{} -- (150:1.5) (0,0) -- (-150:1.5) (1,0) node[ns]{} -- ++(30:1.5) --++(60:1) node[ns]{} (1,0) ++(30:1.5) -- ++(1,0) node[ns]{} (1,0) -- ++(-30:1.5) -- ++(1,0) node[ns]{} (1,0) ++(-30:1.5) -- ++(-60:1) node[ns]{} node[text=black,right=0pt]{$r$};
	\draw [red,very thick] (150:1.5) -- ++(-1,0) node[ns]{} (150:1.5) -- ++(120:1) node[ns]{} node[above=0pt,text=black]{$l$} (-150:1.5) -- ++(-1,0) node[ns]{} (-150:1.5) -- ++(-120:1) node[ns]{};
	\draw [thin,dashed] (0.2,-2) rectangle (-2.6,2.2) (0.8,-2) rectangle (3.6,2.2);
	\draw [blue,postaction={decorate},arrowmark={0.3}{Stealth},postaction={decorate},arrowmark={0.7}{Stealth}] (-1.5,1.8) .. controls (-0.6,0) and (1.6,0) .. (2.5,-1.8);
	\draw [line width=5pt] (0,0) -- (1,0);
	\filldraw [draw=black,fill=white,thick] (150:1.5) circle (0.4cm) (-150:1.5) circle (0.4cm) (1,0) ++(30:1.5) circle (0.4cm) (1,0) ++(-30:1.5) circle (0.4cm);
	\filldraw (-1.5,0) circle (1pt) (170:1.5) circle (1pt) (190:1.5) circle (1pt) (2.5,0) circle (1pt) (1,0) ++(10:1.5) circle (1pt) (1,0) ++(-10:1.5) circle (1pt);
	\node at (-0.1,1.9) {$\mathsf{L}$};
	\node at (1.1,1.9) {$\mathsf{R}$};
	\node at (0,0) [below=0pt]{$i$};
	\node at (1,0) [above=0pt]{$j$};
	\end{tikzpicture}
	\caption{The graph for general connected topology. The white blobs denote generic trees composed of red edges and solid dots.}
	\label{fig:connectM}
\end{figure}

Next, we move on to generic connected topologies, as shown in Figure~\ref{fig:connectM}. The nodes are separated into two sets $\mathsf{L}$ and $\mathsf{R}$, both of which are trees rooted on the node of the tensor current $J_{\hat \varepsilon_1\otimes\varepsilon_i\otimes\varepsilon_j}$. For any two nodes $l\in\mathsf{L}$ and $r\in\mathsf{R}$, there is a unique path connecting them, passing through the roots $i$ and $j$. This path, denoted as $(l,r)$ and schematically shown by the blue curve in Figure~\ref{fig:connectM}, traverses an ordered subset of indices that starts at $l$ and ends at $r$. We find that the coefficient $R_h$ of a monomial with a generic connected topology can be related to a linear combination of the binary BCJ relations in the following way
\begin{equation}\label{eq:connectRh}
  \sum_{\rho\in S_{n-2}}R_h(\sigma)A(\sigma)=\sum_{\rho\in O_{\text{tensor}}}\mathcal{B}(\rho_2\rho_3\cdots\rho_{n-1})\,,
\end{equation}
where $O_{\text{tensor}}$ contains \emph{all} the permutations that respect the relative orderings of indices given by \emph{all} possible pairs $(l,r)$. For instance, we consider the following monomial that appears at six points:
\begin{align}\label{eq:connectexample}
  \varepsilon _2\Cdot p_4 \,\varepsilon _3\Cdot p_4 \,J_{\varepsilon _1\otimes \varepsilon _4\otimes \varepsilon _5}:\qquad\begin{tikzpicture}[baseline={([yshift=-0.7ex]current bounding box.center)}]
    \draw [red,very thick] (150:1) -- (0,0) -- (-150:1);
    \draw [line width=5pt] (0,0) -- (0.5,0);
    \filldraw (0,0) circle (2.5pt) node[above=1pt]{$4$} (0.5,0) circle (2.5pt) node[above=1pt]{$5$} (150:1) circle (2.5pt) node[left=1pt]{$2$} (-150:1) circle (2.5pt) node[left=1pt]{$3$};
  \end{tikzpicture}\,.
\end{align}
In this case, we have $\mathsf{L}=\{2,3,4\}$ and $\mathsf{R}=\{5\}$, which give three $(l,r)$ paths:
\begin{align}\label{eq:6psubsets}
  (l,r)=(2,5)\rightarrow\{2,4,5\}\,,& &(l,r)=(3,5)\rightarrow\{3,4,5\}\,,& &(l,r)=(4,5)\rightarrow\{4,5\}\,.
\end{align}
Among all the permutations of $\{2,3,4,5\}$, only $\{2,3,4,5\}$ and $\{3,2,4,5\}$ are compatible with all the three ordered subsets in~eq.~\eqref{eq:6psubsets}, namely
\begin{align}
  O_{\text{tensor}}=\big\{\{2,3,4,5\},\{3,2,4,5\}\big\}\,.
\end{align}
We can read off the coefficient vector $R_h$ of $\varepsilon _2\Cdot p_4 \,\varepsilon _3\Cdot p_4 \,J_{\varepsilon _1\otimes \varepsilon _4\otimes \varepsilon _5} $ from eq.~\eqref{eq:connectRh}:
\begin{align}
  \sum_{\sigma\in S_3}R_h(\sigma)A(\sigma)=\mathcal{B}(2345)+\mathcal{B}(3245)\,.
\end{align}
which agrees with that in eq.~\eqref{eq:sixBCJNumTen}. We note that $O_{\text{tensor}}$ can be generated by the \emph{ordered splitting} of $\mathsf{L}$ followed by the reverse ordered splitting of $\mathsf{R}$. Such operation has appeared before in different but potentially related studies~\cite{Fu:2017uzt,Teng:2017tbo,Gao:2017dek}.

\subsection{Generalized binary BCJ relations and disconnected topologies}\label{sec:disconnected}
Now we move on to the tensor monomials with disconnected topologies. These tensor monomials contain at least one factor of $\varepsilon_i\Cdot p_1$. The discussion here resembles that for connected topologies. The coefficient vectors of these monomials can be related to the generalized binary BCJ relations, defined as the following object being zero on shell: 
\begin{align}\label{eq:BiLocalBCJ}
 \mathcal{B}(\beta_1\beta_2\cdots\beta_s|\alpha_1\alpha_2\cdots\alpha_r)&\equiv \left(\mathbb{I}-\mathbb{P}_{(\alpha_1\cdots\alpha_{r})}\right)\cdot\left(\mathbb{I}-\mathbb{P}_{(\alpha_1\cdots\alpha_{r-1})}\right)\cdots\left(\mathbb{I}-\mathbb{P}_{(\alpha_1\alpha_2)}\right)\nonumber\\
 &\quad\cdot\sum_{\substack{\rho\in\{\beta_1\cdots\beta_s\} \\ ~\shuffle\{\alpha_1\cdots\alpha_r\}}}\sum_{\rho_i<\rho_{\alpha_1}}s_{\rho_i\alpha_1}A(1,\rho,n)\,,
\end{align}
where $\alpha$ and $\beta$ are two disjoint ordered sets whose union gives $\{2,3\cdots n-1\}$. Suppose $\alpha_1$ is the $(j+1)$-th element in the color ordering $\{1,\rho,n\}$, then the coefficient of $A(1,\rho,n)$ can be written explicitly as
\begin{align}
  \sum_{\rho_i<\rho_{\alpha_1}}s_{\rho_i\alpha_1}=s_{1\rho_2\cdots\rho_j\alpha_1}-s_{1\rho_2\cdots\rho_j}\,.
\end{align} 
In particular, if $\alpha_1=\rho_2$, we have $s_{1\rho_2\cdots\rho_j\alpha_1}=s_{1\alpha_1}$ and $s_{1\rho_2\cdots\rho_j}=0$. Eq.~\eqref{eq:BiLocalBCJ} returns to a binary BCJ relation if $\beta$ is empty.
The generalized binary BCJ relations are obtained from consecutive actions of the operators of the form $(\mathbb{I}-\mathbb{P}_{(\alpha_1\alpha_2\cdots)})$, which double the number of terms after every action. The second line of eq.~\eqref{eq:BiLocalBCJ} contains $\frac{(r+s)!}{r!s!}$ color ordered amplitudes and hence the total number of color ordered amplitudes in $\mathcal{B}(\beta_1\beta_2\cdots\beta_s|\alpha_1\alpha_2\cdots\alpha_r)$ is precisely $2^{r-1}\frac{(r+s)!}{r!s!}$. We can rewrite eq.~\eqref{eq:BiLocalBCJ} as
 \begin{align}
   &\mathcal{B}(\beta_1\beta_2\cdots\beta_s|\alpha_1\alpha_2\cdots\alpha_r)= \nonumber\\
   &\qquad\sum_{i=1}^{r}\sum_{\substack{\gamma\in\{\alpha_{i-1}\cdots\alpha_1\} \\ \shuffle\{\alpha_{i+1}\cdots\alpha_r\}}}\sum_{\substack{\rho\in\{\beta_1\cdots\beta_s\} \\ \shuffle \{\alpha_i,\gamma\}}} (-1)^{i-1}\sum_{\rho_j<\rho_{\alpha_i}}s_{\sigma_j\alpha_i}A(1,\rho,n)\,.
 \end{align}
On the support of momentum conservation, it is also a linear combination of the generalized BCJ relations
\begin{align}\label{BL2BCJ}
  \mathcal{B}(\beta_1\beta_2\cdots\beta_s|\alpha_1\alpha_2\cdots\alpha_r)&=\sum_{i=1}^{r}(-1)^{i-1}\!\sum_{\substack{\rho\in\{\alpha_{i-1}\cdots\alpha_2,\alpha_1\} \\ \shuffle \{\beta_1,\beta_2\cdots\beta_s\}}}\!\mathcal{R}(\rho |\alpha_i\cdots\alpha_r)\,.
\end{align}
For example, at $n=5$, $\mathcal{B}(2|34)$ is given by
\begin{align}\label{eq:fiveBLBCJ}
\mathcal{B}(2|34)&=\left(\mathbb{I}-\mathbb{P}_{(34)}\right)\cdot\left[(s_{123}-s_{12})A(12345)+s_{13}A(13245)+s_{13}A(13425)\right]\nonumber\\
&=(s_{123}-s_{12})A(12345)+s_{13}A(13245)+s_{13}A(13425)\nonumber\\
&\quad-(s_{124}-s_{12})A(12435)+s_{14}A(14235)+s_{14}A(14325)\,,
\end{align}
while the right hand side of eq.~\eqref{BL2BCJ} reads
\begin{align}\label{eq:fiveBCJ}
\mathcal{R}(2|34)&-\mathcal{R}(23|4)-\mathcal{R}(32|4)\nonumber\\
&=(s_{1234}-s_{12})A(12345)+(s_{1234}-s_{123}+s_{13})A(13245)+s_{134}A(13425)\nonumber\\
&\quad-(s_{1234}-s_{123})A(12345)+(s_{124}-s_{12})A(12435)+s_{14}A(14235)\nonumber\\
&\quad-(s_{1324}-s_{132})A(13245)+(s_{134}-s_{13})A(13425)+s_{14}A(14325)\,,
\end{align}
which agrees with eq.~\eqref{eq:fiveBLBCJ} after using the momentum conservation. Another example of eq.~\eqref{BL2BCJ} at $n=6$ is
\begin{align}
\mathcal{B}(23|45)=\mathcal{R}(23|45)-\mathcal{R}(234|5)-\mathcal{R}(243|5)-\mathcal{R}(423|5)\,,
\end{align}
which can also be directly verified. We have checked the validity of eq.~\eqref{BL2BCJ} numerically up to twenty points.

Next, we give the prescription for reading off the $R_h$'s of the tensor monomials with disconnected topologies. Since these tensor monomials contain at least one factor of $\varepsilon_i\Cdot p_1$, their diagrams contain one or more trees that are rooted on a white node.

\begin{figure}[t]
  \centering
  \begin{tikzpicture}[ns/.style={circle,fill=black,inner sep=0,minimum size=5pt},arrowmark/.style 2 args={decoration={markings,mark=at position #1 with \arrow{#2}}}]
    \draw [red,very thick] (0,0) node[ns]{} -- (150:1.5) (0,0) -- (-150:1.5) (1,0) node[ns]{} -- ++(30:1.5) --++(60:1) node[ns]{} (1,0) ++(30:1.5) -- ++(1,0) node[ns]{} (1,0) -- ++(-30:1.5) -- ++(1,0) node[ns]{} (1,0) ++(-30:1.5) -- ++(-60:1) node[ns]{};
    \draw [red,very thick] (150:1.5) -- ++(-1,0) node[ns]{} (150:1.5) -- ++(120:1) node[ns]{} (-150:1.5) -- ++(-1,0) node[ns]{} (-150:1.5) -- ++(-120:1) node[ns]{};
    \draw [thin,dashed] (0.2,-2) rectangle (-2.6,2.2) (0.8,-2) rectangle (3.6,2.2);
    \draw [line width=5pt] (0,0) -- (1,0);
    \filldraw [draw=black,fill=white,thick] (150:1.5) circle (0.4cm) (-150:1.5) circle (0.4cm) (1,0) ++(30:1.5) circle (0.4cm) (1,0) ++(-30:1.5) circle (0.4cm);
    \filldraw (-1.5,0) circle (1pt) (170:1.5) circle (1pt) (190:1.5) circle (1pt) (2.5,0) circle (1pt) (1,0) ++(10:1.5) circle (1pt) (1,0) ++(-10:1.5) circle (1pt);
    \node at (-0.1,1.9) {$\mathsf{L}$};
    \node at (1.1,1.9) {$\mathsf{R}$};
    \node at (0,0) [below=0pt]{$i$};
    \node at (1,0) [below=0pt]{$j$};
    \begin{scope}[xshift=-6cm,yshift=2.1cm,rotate=20]
      \draw [red,very thick] (0,0) -- (1.5,0) -- ++(30:1) (1.5,0) -- ++(-30:1);
      \filldraw [draw=black,fill=white,thick] (1.5,0) circle (0.4cm);
      \filldraw (1.5,0) ++(30:1) circle (2.5pt) node[right=0pt]{$t_\ell$} (1.5,0) ++(-30:1) circle (2.5pt);
    \end{scope}
    \begin{scope}[xshift=-6cm,yshift=2.1cm,rotate=-20]
      \draw [red,very thick] (0,0) -- (1.5,0) -- ++(30:1) (1.5,0) -- ++(-30:1);
      \filldraw [draw=black,fill=white,thick] (0,0) circle (2.5pt) node [below=0pt]{$t_1$} (1.5,0) circle (0.4cm);
      \filldraw (1.5,0) ++(30:1) circle (2.5pt) (1.5,0) ++(-30:1) circle (2.5pt);
    \end{scope}
    \begin{scope}[xshift=-6cm,yshift=2.1cm]
      \filldraw (1,0) circle (1pt) (10:1) circle (1pt) (-10:1) circle (1pt);
      \draw [blue,postaction={decorate},arrowmark={0.3}{Stealth},postaction={decorate},arrowmark={0.7}{Stealth}] (0,0.3) .. controls (1, 0.8) and (1,0.8) .. (1.7,1.3);
    \end{scope}
    \begin{scope}[xshift=-6cm,yshift=-0.9cm,rotate=20]
      \draw [red,very thick] (0,0) -- (1.5,0) -- ++(30:1) (1.5,0) -- ++(-30:1);
      \filldraw [draw=black,fill=white,thick] (1.5,0) circle (0.4cm);
      \filldraw (1.5,0) ++(30:1) circle (2.5pt) (1.5,0) ++(-30:1) circle (2.5pt);
    \end{scope}
    \begin{scope}[xshift=-6cm,yshift=-0.9cm,rotate=-20]
      \draw [red,very thick] (0,0) -- (1.5,0) -- ++(30:1) (1.5,0) -- ++(-30:1);
      \filldraw [draw=black,fill=white,thick] (0,0) circle (2.5pt) (1.5,0) circle (0.4cm);
     \filldraw (1.5,0) ++(30:1) circle (2.5pt) (1.5,0) ++(-30:1) circle (2.5pt);
    \end{scope}
    \begin{scope}[xshift=-6cm,yshift=-0.9cm]
      \filldraw (1,0) circle (1pt) (10:1) circle (1pt) (-10:1) circle (1pt);
    \end{scope}
    \begin{scope}[xshift=-6cm,yshift=-0.9cm]
      \draw [thin,dashed]  (2.8,1.5) rectangle (-0.5,4.5);
      \node at (-0.2,4.2) {$\mathsf{T}_1$};
      \draw [thin,dashed]  (2.8,-1.5) rectangle (-0.5,1.5);
      \node at (-0.2,1.1) {$\mathsf{T}_2$};
    \end{scope}
        \begin{scope}[xshift=-6cm,yshift=-2.9cm]
      \filldraw (1,0) circle (1pt) (10:1) circle (1pt) (-10:1) circle (1pt);
    \end{scope}
  \end{tikzpicture}
  \caption{The graph for general disconnected topology.}
  \label{fig:disconnectM}
\end{figure}

The simplest disconnected topology contains two chains, one of which starts at a white node as shown below
\begin{align}\label{eq:doublechain}
  \begin{tikzpicture}[scale=1.0]
    \draw [red,very thick] (0.5,0) -- (1.5,0) (0,0) -- (-1,0) (2.5,0) -- (3.5,0) (-2,0) -- (-3,0);
    \draw [red,very thick,dashed] (-1,0) -- (-2,0) (1.5,0) -- (2.5,0);
    \filldraw [black] (0,0) circle (2.5pt) node [above=0pt]{$\alpha_{i\vphantom{+}}$} (0.5,0) circle (2.5pt) node [above=0pt]{$~~\alpha_{i+1}$} (1.5,0) circle (2.5pt) node[above=0pt]{$\alpha_{i+2}$} (2.5,0) circle (2.5pt) node[above=0pt]{$\alpha_{r-1}$} (3.5,0) circle (2.5pt) node[above=0pt]{$\alpha_{r}$} (-1,0) circle (2.5pt) node[above=0pt]{$\alpha_{i-1}$} (-2,0) circle (2.5pt) node[above=0pt]{$\alpha_2$} (-3,0) circle (2.5pt) node[above=0pt]{$\alpha_1$};
    \draw [line width=5pt] (0,0) -- (0.5,0);
    \begin{scope}[xshift=-7cm]
      \draw [red,very thick] (0,0) -- (1,0) (2,0) -- (3,0);
      \draw [red,dashed,very thick] (1,0) -- (2,0);
      \filldraw [draw=black,thick,fill=white] (0,0) circle (2.5pt) node[above=0pt]{$\beta_1$};
      \filldraw [black] (1,0) circle (2.5pt) node[above=0pt]{$\beta_2$} (2,0) circle (2.5pt) node[above=0pt]{$\beta_{s-1}$} (3,0) circle (2.5pt) node[above=0pt]{$\beta_s$};
    \end{scope}
  \end{tikzpicture}\,.
\end{align}
This diagram comes from the tensor monomial
\begin{align*}
  \big[\varepsilon_{\beta_s}\Cdot p_{\beta_{s-1}}\cdots\varepsilon_{\beta_2}\Cdot p_{\beta_1}\varepsilon_{\beta_1}\Cdot p_1\big] \times \big[\varepsilon_{\alpha_1}\Cdot p_{\alpha_2} \cdots \varepsilon_{\alpha_{i-1}}\Cdot p_{\alpha_i} J_{\hat \varepsilon_1\otimes \varepsilon_{\alpha_i}\otimes \varepsilon_{\alpha_{i+1}}} \, \varepsilon_{\alpha_{i+2}}\Cdot p_{\alpha_{i+1}} \cdots \varepsilon_{\alpha_{r}} \Cdot p_{\alpha_{r-1}}\big]\,.
\end{align*}
We find that the coefficient $R_h$ of this monomial is related to a single generalized binary BCJ relation as follows
\begin{align}\label{eq:dcrule}
\sum_{\sigma\in S_{n-2}}R_h(\sigma)A(\sigma)=\mathcal{B}(\beta_1\beta_2\cdots\beta_s|\alpha_1\alpha_2\cdots\alpha_r)\,.
\end{align}
We consider the monomial $\varepsilon_2\Cdot p_1\,J_{\hat \varepsilon_1\otimes\varepsilon_3\otimes\varepsilon_4}$ as an example, which appears at five points. The diagram for this monomial is
\begin{align}\label{eq:discon1}
  \varepsilon_2\Cdot p_1\,J_{\hat \varepsilon_1\otimes\varepsilon_3\otimes\varepsilon_4}:\qquad
  \begin{tikzpicture}[baseline={([yshift=-0.8ex]current bounding box.center)}]
    \draw [line width=5pt] (0,0) -- (0.5,0);
    \filldraw [black] (0,0) circle (2.5pt) node[above=0pt]{$3$} (0.5,0) circle (2.5pt) node[above=0pt]{$4$};
    \filldraw [draw=black,fill=white,thick] (-1,0) circle (2.5pt) node[above=0pt]{$2$};
  \end{tikzpicture}\,.
\end{align}
The coefficient vector of this monomial is $R_3$ in eq.~\eqref{eq:Rh5pt}, which agrees with the relation below following from eq.~\eqref{eq:dcrule}
\begin{align}
  R_h\text{ ~\,of~\,\,}\varepsilon_2\Cdot p_1 J_{\hat \varepsilon_1\otimes\varepsilon_3\otimes\varepsilon_4} :\qquad\sum_{\sigma\in S_3}R_h(\sigma)A(\sigma)=\mathcal{B}(2|34)\,.
\end{align}

For a generic disconnected topology as shown in Figure~\ref{fig:disconnectM}, where the trees rooted on white nodes are labeled as $\mathsf{T}_1$, $\mathsf{T}_2$, etc, we can relate the coefficient vector $R_h$ of such a topology to a linear combination of the generalized BCJ relations as follows
\begin{align}\label{eq:disconRh}
  \sum_{\rho}R_h(\rho)A(\rho)=\sum_{\beta\in O(\mathsf{T}_1,\mathsf{T}_2\cdots)}\sum_{\alpha\in O_{\text{tensor}}}\mathcal{B}(\beta_1\beta_2\cdots\beta_s|\alpha_1\alpha_2\cdots\alpha_s)\,,
\end{align}
where the summations are taken over two sets of color orderings. While $O_{\text{tensor}}$ is constructed from the tensor part of this monomial the same way as in eq.~\eqref{eq:connectRh}, the set $O(\mathsf{T}_1,\mathsf{T}_2\cdots)$ is constructed similarly from the trees rooted on white nodes. Given such a tree, say, $\mathsf{T}_1$, a path from the root $t_1$ to another node $t_\ell$ (the blue path in Figure~\ref{fig:disconnectM}) defines an ordered subset of the nodes along this path. Then $O(\mathsf{T}_1,\mathsf{T}_2\cdots)$ contains all the permutations that are compatible with all such ordered subsets. {Like the case for connected topologies, ordered splitting~\cite{Fu:2017uzt,Teng:2017tbo,Gao:2017dek} can be used to generate this set.}

\begin{figure}[t]
  \centering
  \subfloat[][$\varepsilon_2\Cdot p_1\varepsilon_3\Cdot p_1\varepsilon_4\Cdot p_1 J_{\hat \varepsilon_1\otimes\varepsilon_5\otimes\varepsilon_6}$]{\begin{tikzpicture}
      \tikzstyle{every node}=[font=\small]
      \draw [line width=5pt] (0,0) -- (0.5,0);
      \filldraw (0,0) circle (2.5pt) node[above=0pt]{$5$} (0.5,0) circle (2.5pt) node[above=0pt]{$6$};
      \filldraw [draw=black,fill=white,thick] (-0.75,0) circle (2.5pt) node[left=0pt]{$3$} (-0.75,0.75) circle (2.5pt) node[left=0pt]{$2$} (-0.75,-0.75) circle (2.5pt) node[left=0pt]{$4$};
      \node at (2,0) {};
      \node at (-2,0) {};
    \end{tikzpicture}\label{fig:7pa}
    }\qquad
    \subfloat[][$\varepsilon_2\Cdot p_1\varepsilon_3\Cdot p_2\varepsilon_4\Cdot p_1 J_{\hat \varepsilon_1\otimes\varepsilon_5\otimes\varepsilon_6}$]{\begin{tikzpicture}
	\tikzstyle{every node}=[font=\small]
	\draw [line width=5pt] (0,0) -- (0.5,0);
        \draw [red,very thick] (-0.5,0.5) -- (-1.5,0.5);
        \filldraw (0,0) circle (2.5pt) node[above=0pt]{$5$} (0.5,0) circle (2.5pt) node[above=0pt]{$6$} (-0.5,0.5) circle (2.5pt) node[above=0pt]{$3$};
        \filldraw [draw=black,fill=white,thick] (-1.5,0.5) circle (2.5pt) node[above=0pt]{$2$} (-1.5,-0.5) circle (2.5pt) node[below=0pt]{$4$};
        \node at (2,0) {};
        \node at (-2,0) {};
      \end{tikzpicture}\label{fig:7pb}
    }\qquad
    \subfloat[][$\varepsilon_2\Cdot p_1\varepsilon_3\Cdot p_2\varepsilon_4\Cdot p_2 J_{\hat \varepsilon_1\otimes\varepsilon_5\otimes\varepsilon_6}$]{\begin{tikzpicture}
	\tikzstyle{every node}=[font=\small]
	\draw [line width=5pt] (0,0) -- (0.5,0);
        \draw [red,very thick] (-1.5,0) -- ++(20:1) (-1.5,0) -- ++(-20:1);
        \filldraw (0,0) circle (2.5pt) node[above=0pt]{$5$} (0.5,0) circle (2.5pt) node[above=0pt]{$6$} (-1.5,0) ++(20:1) circle (2.5pt) node[above=0pt]{$3$} (-1.5,0) ++(-20:1) circle (2.5pt) node[below=0pt]{$4$};
        \filldraw [draw=black,fill=white,thick] (-1.5,0) circle (2.5pt) node[above=0pt]{$2$};
        \node at (2,0) {};
        \node at (-2,0) {};
      \end{tikzpicture}\label{fig:7pc}
    }
    \caption{Typical monomials of disconnected topologies at seven points.}
    \label{fig:7pmon}
\end{figure}
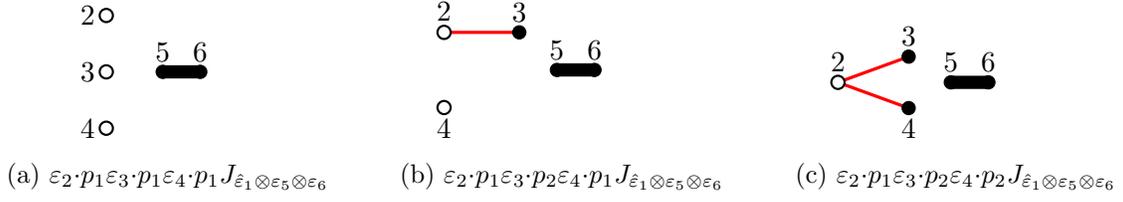

We give three examples shown in Figure~\ref{fig:7pmon}. They are all monomials with disconnected topologies that appear at seven points.
They have the same trivial tensor part such that $O_{\text{tensor}}=\{\{5,6\}\}$. For the monomial of Figure~\ref{fig:7pa}, there are no constraints on the ordering of the three white nodes, and hence eq.~\eqref{eq:dcrule} demands
\begin{align}
\text{Figure~\ref{fig:7pa}:}\qquad\sum_{\sigma\in S_5}R_h(\sigma)A(\sigma)&=\mathcal{B}(234|56)+\mathcal{B}(243|56)+\mathcal{B}(324|56)\nonumber\\
&\quad+\mathcal{B}(342|56)+\mathcal{B}(423|56)+\mathcal{B}(432|56)\,.
\end{align}
In Figure~\ref{fig:7pb}, a tree stems from one of the white nodes and the order $\{2,3\}$ has to be respected. In this case, eq.~\eqref{eq:dcrule} reads
\begin{align}
\text{Figure~\ref{fig:7pb}:}\qquad\sum_{\sigma\in S_5}R_h(\sigma)A(\sigma)&=\mathcal{B}(234|56)+\mathcal{B}(243|56)+\mathcal{B}(423|56)\,.
\end{align}
Finally, in Figure~\ref{fig:7pc}, there are two paths constructed from the part containing the white nodes, which define the orders $\{2,3\}$ and $\{2,4\}$. Eq.~\eqref{eq:dcrule} reads
\begin{align}
\text{Figure~\ref{fig:7pc}:}\qquad\sum_{\sigma\in S_5}R_h(\sigma)A(\sigma)&=\mathcal{B}(234|56)+\mathcal{B}(243|56)\,.
\end{align}
All the three results agree with the fusion product calculation at seven points.

\section{Conclusion and Outlook}
In this paper, we introduce a framework that makes use of tensor currents and fusion products to construct BCJ numerators in the DDM basis. This framework constitutes a particular realization of the kinematic Lie algebra that underlies color-kinematics duality, and that was first studied by Monteiro and O'Connell in the self-dual sector of Yang-Mill theory~\cite{Monteiro:2011pc}. We restrict the discussion in this paper to the bi-scalar NMHV sector~\eqref{eq:ss}, in which the kinematic numerators contain an overall factor $ \varepsilon_1\Cdot  \varepsilon_n$ and there is at most one other factor of $ \varepsilon_i\Cdot \varepsilon_j$  (the polarization power is two). Given certain assumptions, we find a kinematic algebra without free parameters in this sector.  It is defined in terms of  the fusion products of the tensor and vector currents. The uncovered algebra has a simple structure that allows us to write down an all-multiplicity closed formula for the BCJ numerators in this bi-scalar NMHV sector. We note that focusing on the bi-scalar NMHV sector is not a severe limitation, since this sector contains the full information of the NMHV numerators.  Indeed, pure-gluon NMHV numerators can be obtained by superposing bi-scalar numerators following the dimensional-oxidation prescription of ref.~\cite{Chiodaroli:2017ngp}.

In addition to interpreting the vector/tensor currents and fusion products as elements of the kinematic Lie algebra, they can be viewed as the ingredients of a Berends-Giele recursion for off-shell BCJ numerators.  The vector and tensor currents can then be formally viewed as certain off-shell fields that are being recursively constructed. Previous work have systematically analyzed the construction of BCJ numerators through Berends-Giele recursion~\cite{Mafra:2011kj,Mafra:2015vca,Garozzo:2018uzj}, however in that work a non-linear gauge transformation was needed to make the off-shell numerators obey color-kinematics duality. Such gauge transformations were constructed through an iterative process that generated new formulas at each step. For the kinematic algebra constructed in this work, we directly land on a ``BCJ gauge''~\cite{Lee:2015upy} through  recursive application of the closed set of fusion rules. It would be interesting to understand if our construction can be lifted to off-shell fields using an Lagrangian description (see the Lagrangian construction of ref.~\cite{Bern:2010yg}); we leave this to future work.   

The BCJ numerators obtained in our framework inherit a large amount of generalized gauge freedom that originate from the unphysical degrees of freedom that are associated to both the vector and tensor currents. In particular, as is well-known in the standard treatment of gauge theory,  off-shell vectors carry unphysical scalar degrees of freedom. On the other hand, the tensor currents in our framework obey relations inherited from the Clifford algebra, allowing us to decompose them into gauge-dependent vectors and irreducible tensors.
The freedom in choosing these irreducible tensors is captured by a set of free parameters, which also appears naturally in the induced gauge-dependent vectors. While the irreducible tensors are unphysical and get killed by a null space when multiplying the BCJ numerators with the propagator matrix, the gauge-dependent vector degrees of freedom become part of the generalized gauge freedom, quantitatively described by the set of free parameters.
One can further constrain such freedom by demanding that the numerators are manifestly crossing symmetric. To aid the analysis of these free parameters we introduced a diagrammatic representation of the tensor monomials, where different topologies correspond to independent tensor monomials.
Using these diagrams one can also straightforwardly show that the tensor monomials are in the null space of the propagator matrix. Given a tensor monomial, we present a prescription to write down the relevant generalized BCJ relations, in which the coefficient vector of the color-ordered amplitudes are exactly that of the given tensor monomial.

To extend the kinematic algebra to the full NMHV sector of Yang-Mills theory, one needs use the fusion product to define commutators between all pairs of vector and tensor currents. The commutators can be used to generate BCJ numerators that automatically have the correct crossing symmetry (including the first leg) and make manifest the kinematic Jacobi identity, rather than implicitly assuming the Jacobi identity as we do this work. 
We will explore these details in a forthcoming paper, where we present a kinematic algebra for the full NMHV sector of Yang-Mills theory. 
 
In the sectors with even higher polarization powers, e.g.~$ (\varepsilon_i\Cdot \varepsilon_j)^{k\ge3}$ needed for computing N$^{k-1}$MHV amplitudes,  higher-rank tensor currents are likely required both for a consistent algebra and to account for all of the generalized gauge freedom. It starts becoming computationally challenging to find the fusion rules involving these higher-rank tensor currents, since in principle one needs to check consistency of the construction beyond multiplicity-eight amplitudes. As a possible alternative to working with high-multiplicity amplitudes between physical states, it would be interesting if one can carry out the algebraic construction by analyzing three- and four-point amplitudes between all physical and unphysical states. One may suspect this should be possible, given that  the fusion product encodes cubic interactions, and the kinematic Jacobi identity is first relevant at the four-point level. 

Finally, an interesting open question that we have not addressed in this work, is to better understand the significance of a hidden kinematic Lie algebra in Yang-Mills theory. The existence of a Lie algebra usually implies a symmetry, but in the case of the kinematic Lie algebra one should expect that not all of the presumed generators can correspond to symmetries. The reason is that the Lie algebra is only realized for the kinematic numerators, and in a scattering amplitude the numerators are also weighed by the momentum-dependent propagators, which may interfere with the symmetry transformations. Indeed, a similar phenomenon occurs in the case of spontaneously broken symmetries, where the Lie algebra of the broken symmetry directions can still be realized~\cite{Chiodaroli:2015rdg}. Clearly this question about symmetries is important and deserves further investigations.

\acknowledgments
We thank M. Chiodaroli, Y. Du, B. Feng, A. Ochirov, O. Schlotterer and E. Yuan for interesting discussions related to this work. We also thank G. Mogull and G. K\"{a}lin for sharing of some useful code. T.W. thanks J. Plefka and G.C. thanks A. Brandhuber, R. Monteiro, G. Travaglini, C. Wen and C. White for interesting discussions or collaborations on related work. The research is supported by the Swedish Research Council under grant 621-2014-5722, the Knut and Alice Wallenberg Foundation under grant KAW 2013.0235, and the Ragnar S\"{o}derberg Foundation (Swedish Foundations' Starting Grant). G.C. is also supported by the Science and Technology Facilities Council (STFC) Consolidated Grant ST/P000754/1 ``String theory, gauge theory and duality'' and NSF of China Grant under Contract 11405084. T.W. is also supported by the China Scholarship Council for support (File No. 201706190098) and the German Research Foundation through grant PL457/3-1.

\bibliographystyle{JHEP}
\bibliography{ScatEq}

\end{document}